%% file: squeezing_paper_reordered_nohighlit.tex
\documentclass[twocolumn,aps,pra,superscriptaddress]{revtex4-2}

\usepackage{graphicx}
\usepackage{dcolumn}
\usepackage{bm}
\usepackage{subcaption}
\usepackage{xcolor}
\newcommand{\rev}[1]{#1}
\newcommand{\bigrev}[1]{#1}

\DeclareCaptionJustification{fulljustify}{\leftskip=0pt \rightskip=0pt \parfillskip=0pt plus 1fil}
\usepackage{amsmath, amssymb, physics}
\usepackage{siunitx}
\usepackage{tikz}
\usepackage{braket}
\usepackage{dsfont}
\usepackage{tabularx}
\usepackage{babel}
\usepackage{hyperref}

\renewcommand{\selectlanguage}[1]{} 

\newcommand{\textsubcap}[1]{#1}

\begin{document}
\preprint{APS/123-QED}

\setcounter{secnumdepth}{3}

\title{\textbf{Practical Limits to Single-Mode Vacuum Squeezing with a SNAIL Parametric Amplifier}}




\author{Theodore Shaw}
\email{Contact author: shaw.th@utexas.edu}
\affiliation{Department of Electrical and Computer Engineering, University of Texas at Austin, Austin, TX, USA}%

\author{Debsuvra Mukhopadhyay}
\affiliation{Department of Physics, Northwestern University, Evanston, IL, USA}

\author{Zhuoqun Hao}
\affiliation{Department of Electrical and Computer Engineering, University of Texas at Austin, Austin, TX, USA}

\author{Josiah Cochran}
\affiliation{Department of Electrical and Computer Engineering, University of Texas at Austin, Austin, TX, USA}

\author{Haley M. Cole}
\affiliation{Department of Electrical and Computer Engineering, University of Texas at Austin, Austin, TX, USA}

\author{Archana Kamal}
\affiliation{Department of Physics, Northwestern University, Evanston, IL, USA}

\author{Shyam Shankar}
\email{Contact author: shyam.shankar@utexas.edu}
\affiliation{Department of Electrical and Computer Engineering, University of Texas at Austin, Austin, TX, USA}







\date{\today}

\begin{abstract}
We characterize single-mode vacuum squeezing generated by a SNAIL Parametric Amplifier (SPA) operated under conditions representative of practical sensing and qubit-readout experiments. Motivated by prior expectations that Kerr-induced distortion limits squeezing in degenerate parametric amplifiers, we studied squeezing as a function of external flux and pump power to explore operating points where Kerr nonlinearity is theoretically minimized. We find that for squeezing operation at a fixed frequency, Kerr varied by about a factor of two and the achievable squeezing showed no significant dependence on Kerr. 
Rather than Kerr, the squeezing is dominated by internal resonator loss and insertion loss in the microwave chain as confirmed by theoretical modeling. These results indicate that, in practical SPAs, reducing loss, rather than suppressing Kerr, is the primary route to improved squeezing performance.

\end{abstract}

\maketitle


\section{\label{sec:intro}Introduction}
A squeezed vacuum state is a valuable resource for enhancing the sensitivity of quantum measurements. By replacing the input state of a system with squeezed vacuum, the noise in one quadrature can be reduced, thereby improving the signal-to-noise ratio of a phase-sensitive measurement \cite{walls_quantum_2025}. Squeezed states have been used for this purpose in a variety of quantum sensing experiments at optical and microwave frequencies~\cite{the_ligo_scientific_collaboration_gravitational_2011, bienfait_magnetic_2017, malnou_squeezed_2019,  backes_quantum_2021}, as well as for improving the fidelity of qubit readout in superconducting quantum devices~\cite{barzanjeh_dispersive_2014, didier_fast_2015, didierHeisenbergLimitedQubitReadOut2015, eddins_stroboscopic_2018, liu_noise_2022}. In microwave frequency experiments, squeezed vacuum states are typically produced by parametric amplifiers based on Josephson junctions or kinetic inductance \cite{movshovich_observation_1990, lvovsky_continuous-variable_2009, eichler_experimental_2011, mallet_quantum_2011, zhou_high-gain_2014, boutinEffectHigherorderNonlinearities2017, malnouOptimalOperationJosephson2018, perelshtein_broadband_2022, qiu_broadband_2023, vaartjesStrongMicrowaveSqueezing2024}. 
Recent advances have seen single-mode squeezing of around 10 dB~\cite{qiu_broadband_2023} and even under challenging conditions such as high magnetic fields and elevated temperatures \cite{vaartjesStrongMicrowaveSqueezing2024}. However, in practical sensing applications, the usable squeezing has typically been closer to 4 dB, as demonstrated in axion dark matter searches \cite{malnou_squeezed_2019, backes_quantum_2021}. The usable squeezing is limited by losses in the signal path between the squeezer and the measurement device, but also has been shown to be limited by distortions introduced by parasitic nonlinearities in the squeezer\cite{boutinEffectHigherorderNonlinearities2017,bienfait_magnetic_2017}.

In this article, we characterize the microwave single-mode squeezed state produced by a SNAIL Parametric Amplifier (SPA) \cite{spa1}, where SNAIL refers to the Superconducting Nonlinear Asymmetric Inductive eLement (SNAIL)~\cite{snail}. SPAs are three-wave-mixing amplifiers with \textit{in situ} tunability that enables suppression or elimination of the parasitic four-wave mixing Kerr nonlinearity, which otherwise causes distortion. Recent results have shown that SPAs~\cite{spa1, spa2} and closely related RF-SQUID amplifiers~\cite{sivakJosephsonArrayMode2020, snakeamp} can achieve record compression power for a resonant parametric amplifier due to their low Kerr nonlinearity. In this work, we operate the SPA under conditions representative of a practical sensing or qubit readout experiment and achieve squeezing of about 2 dB. By studying the squeezing as a function of operating parameters and combining experimental results with theoretical analysis, we find that the squeezing level is largely insensitive to Kerr and is instead limited by loss.

\section{SNAIL Parametric Amplifier as a Squeezer}
The unitary operator that generates squeezing in a harmonic oscillator with annihilation operator $\hat{a}$ is
\begin{equation}\label{eq:squeezingOperator}
    \hat{S}(z) = \exp\!\left[\frac{1}{2}\left(z^*\hat{a}^2 - z\hat{a}^{\dagger 2}\right)\right],
\end{equation}
where $z$ is the squeezing parameter~\cite{gerry_introductory_2008}. 
In superconducting quantum circuits, this transformation is realized exactly by the ideal degenerate parametric amplifier (DPA), and approximately realized by the broad class of superconducting parametric amplifiers; when such an amplifier is operated as a phase-sensitive or phase-preserving amplifier, its gain stems from applying the squeezing unitary to the incident microwave field.

The SNAIL parametric amplifier is a DPA whose nonlinearity can be tuned \textit{in situ} via magnetic flux applied to the SNAIL, enabling strong three-wave mixing that generates the desired squeezing while simultaneously suppressing unwanted higher‑order nonlinearities.
The SPA Hamiltonian can be written as
\begin{equation}
    \hat{H}/\hbar = \omega_0 \hat{a}^\dagger \hat{a} + g_3(\hat{a}+\hat{a}^\dagger)^3 + g_4(\hat{a}+\hat{a}^\dagger)^4 + \cdots,
\end{equation}
where $\omega_0$ is the resonance frequency and $g_i$ denote the $i$-th order nonlinear coefficients arising from the SNAIL potential~\cite{spa1}. Both $\omega_0$ and $g_i$ depend on device design and are tunable with external flux $\Phi_{\mathrm{ext}}$. Crucially, the fourth-order nonlinearity $g_4$ vanishes near $\Phi_{\mathrm{ext}}\approx0.4\Phi_0$~\cite{snail, spa1}, potentially allowing  parametric amplification and squeezing from an almost ideal three-wave-mixing DPA. However, as shown in Refs.~\cite{spa1,spa2}, higher-order Hamiltonian corrections and pump‑induced effects complicate this ideal. 

In practical applications, the squeezing frequency $\omega_s$ is fixed by the downstream system, such as a qubit readout cavity~\cite{eddins_stroboscopic_2018} or the probe frequency in a quantum sensing experiment \cite{bienfait_magnetic_2017, malnou_squeezed_2019, backes_quantum_2021}. For a three-wave-mixing DPA, $\omega_s$ is set by the pump frequency $\omega_p$ such that $\omega_s = \omega_p/2$. Assuming the pump is near twice the resonance frequency $\omega_0$, the SPA effective Hamiltonian in a frame rotating at the pump frequency becomes
\begin{equation}
    \hat{H}_{\mathrm{eff}}/\hbar = \Delta_\mathrm{eff} \hat{a}^\dagger \hat{a} + \frac{g_\mathrm{eff}}{2}(\hat{a}^2 + \hat{a}^{\dagger 2}) + \frac{K}{2}\hat{a}^{\dagger 2}\hat{a}^2.
\end{equation}
The parameters $\Delta_\mathrm{eff}$, $g_\mathrm{eff}$ and $K$ depend on all the nonlinearities of $\hat{H}$ as well as the pumping conditions. Analytical expressions for these parameters were derived in Refs.~\cite{spa1, spa2}. To lowest order: (1) the effective detuning is $\Delta_\mathrm{eff} = \Delta + \frac{8}{9}K n_p$ where $\Delta = \omega_0 - \omega_p/2$ is the pump-resonator detuning, $n_p = |\alpha_p|^2$ is the pump-induced intra-resonator photon number, and $\alpha_p$ is the corresponding amplitude; (2) the squeezing rate is $g_\mathrm{eff} = 4g_3 \sqrt{n_p}$; (3) the Kerr nonlinearity $K$, capturing residual four-wave mixing, depends on $g_3$, $g_4$, $\omega_0$, and the pump photon number $n_p$, and to lowest order is~\cite{spa2}

\rev{
\begin{equation}\label{eq:kerr}
    K(n_p) = 12\!\left(g_4 - 5\frac{g_3^2}{\omega_0}\right) + O(n_p) = 12g_4^* + O(n_p),
\end{equation}
where, following Ref.~\cite{spa2}, $g_4^*\equiv g_4 - 5g_3^2/\omega_0$ is the four-wave mixing coefficient obtained at lowest order in harmonic balance.} In practice, however, directly predicting $K(n_p)$ from perturbation theory is unreliable; instead, we adopt an experimentally motivated approach and estimate $|K(n_p)|$ using intermodulation‑distortion (IMD) measurements.

The second term of $\hat{H}_\mathrm{eff}$ generates the squeezing interaction whose dynamics correspond to the unitary in Eq.~(\ref{eq:squeezingOperator}), with strength $g_{\mathrm{eff}}$ set \textit{in situ} by the pump power $P_p\propto n_p$. The effective detuning $\Delta_\mathrm{eff}$ is straightforward to set to zero experimentally, even for fixed $\omega_p/2$, by adjusting $\Phi_{\mathrm{ext}}$ to change $\omega_0$ and thus $\Delta$. By contrast, the Kerr term $K$ is parasitic and cannot be tuned independently of the other two terms of $\hat{H}_\mathrm{eff}$, raising the question of whether its effects can be minimized in practice to realize the squeezing unitary with high fidelity.

To understand how $\hat{H}_\mathrm{eff}$ affects traveling fields incident on the SPA, we use the quantum Langevin equation and input–output theory. The SPA resonator is assumed to be coupled to a transmission line with external coupling rate $\kappa_\mathrm{ext}$. Neglecting internal loss for the moment, the total loss rate $\kappa = \kappa_\mathrm{ext}$. In this situation, the phase-preserving power gain is obtained via semi-classical harmonic balance analysis, previously described in Refs.~\cite{spa1,spa2} and generalized in Appendix~\ref{app:spa-squeezing}. The reflection gain for an incident signal at frequency $\omega_p/2$ is 
\begin{equation}
    G_{\mathrm{IL}} = 1 + \frac{\kappa^2g_\mathrm{eff}^2}{\left(\Delta_\mathrm{eff}^2+(\kappa/2)^2-g_\mathrm{eff}^2\right)}.\label{eq:gain}
\end{equation}
Here $G_{\mathrm{IL}}$ is the power gain in reflection referenced to unity transmission, which differs from the common experimental definition where the gain is referenced to the transmission with the SPA pump turned off. As discussed in Appendix~\ref{app:spa-squeezing}, this distinction becomes important in the presence of loss. At $\omega_p/2$, amplification is phase-sensitive, with the phase of the amplified quadrature $\theta$ set by the phase of the pump drive. The amplitude gain of the amplified quadrature is $\sqrt{G_{\mathrm{IL}}}+\sqrt{G_{\mathrm{IL}}-1}$, while the amplitude gain of the squeezed quadrature is $\sqrt{G_{\mathrm{IL}}}-\sqrt{G_{\mathrm{IL}}-1}$.

From Eq.~\ref{eq:gain}, the gain increases with $n_p$ as $g_\mathrm{eff}$ approaches the parametric instability threshold $\sqrt{\Delta_\mathrm{eff}^2+(\kappa/2)^2}$.  The corresponding vacuum squeezing, with no input, is calculated in Appendix~\ref{app:spa-squeezing}. In the ideal, lossless limit, the squeezed-quadrature variance of the output field is, in photon units,
\begin{equation}
    \mathcal{S}_{\text{min}}
    = \frac{1}{2}
      \left(\sqrt{G_{\mathrm{IL}}} - \sqrt{G_{\mathrm{IL}}-1}\right)^{2}.\label{eq:SminGain}
\end{equation}
Thus, $\mathcal{S}_{\text{min}}$ is not independently tunable from $G_{\mathrm{IL}}$. Instead amplification and squeezing from a parametric amplifier are inextricably linked, and $\mathcal{S}_{\text{min}}$ decreases monotonically to zero as $G_{\mathrm{IL}}\rightarrow\infty$.

Within this linearized description, the explicit role of $K$ is to induce a pump‑dependent Stark shift, causing $\Delta_{\mathrm{eff}}\neq\Delta$ and limiting the maximum achievable gain for fixed $\Delta$ and thus the minimum $\mathcal{S}_{\text{min}}$.  However, if pump parameters ($\Delta$, $n_p$) are adjusted to reach the desired gain, then the squeezed‑quadrature variance is insensitive to $K$ to leading order.

In practice, losses and added noise are always present and arise from two sources: (1) internal dissipation of the SPA resonator, modeled by an internal loss rate $\kappa_\mathrm{int}$ giving $\kappa=\kappa_\mathrm{ext}+\kappa_\mathrm{int}$; and (2) attenuation and added noise of the measurement chain, modeled by an efficiency $\eta < 1$. As shown in Appendix~\ref{app:spa-squeezing}, internal loss destroys the one‑to‑one relation between squeezing and gain given by Eq.~\ref{eq:SminGain}. Nevertheless, when $\Delta_\mathrm{eff}=0$, the observed squeezing obeys
\begin{align}
    \mathcal S_{\mathrm{obs}}=\frac{1}{2}-\frac{(\eta\kappa_{\mathrm{ext}})\,g_{\text{eff}}}{(g_{\text{eff}}+\kappa/2)^2}.
\end{align}
The full expression for general $\Delta_\mathrm{eff}$ is given in Appendix~\ref{app:spa-squeezing}, Eq.~\ref{eq:Sobs}. Importantly, for $\Delta_\mathrm{eff}=0$, $\mathcal S_{\mathrm{obs}}$ is independent of $K$. Though this theory does not capture the Kerr-induced distortion outlined by earlier theoretical work~\cite{boutinEffectHigherorderNonlinearities2017}, this work indicates that in state-of-the-art SPAs, $K/\kappa$ is low enough that squeezing is dominated by loss.


In the following sections, we experimentally investigate how $\mathcal S_{\mathrm{obs}}$ depends on $n_p$ and $\Delta$, under the practical constraint that the squeezing frequency is fixed.



\section{\label{sec:squeezing}Squeezing Measurement}

\begin{figure}
\centering

\begin{subfigure}{0.48\textwidth}
    \stepcounter{subfigure}
    \begin{tikzpicture}
        \node[inner sep=0pt] (img) {\includegraphics{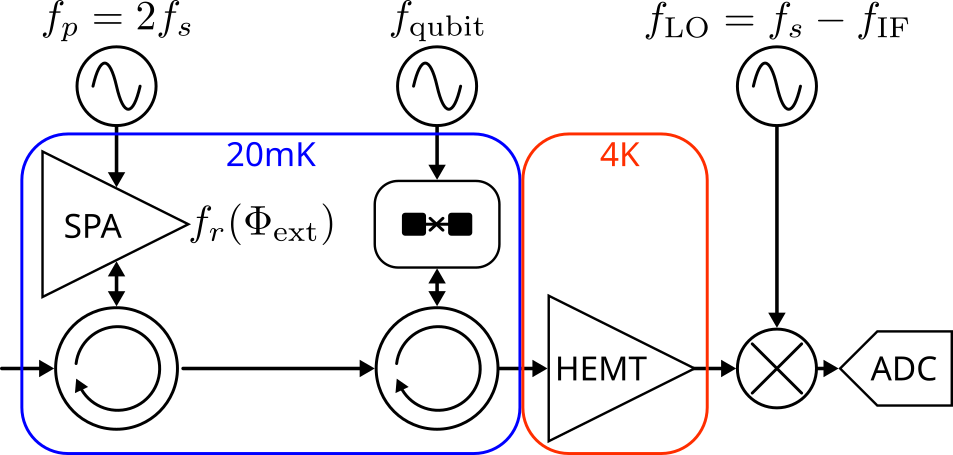}};
        \node[anchor=north west, fill=white, opacity=0.7, text opacity=1, font=\bfseries, xshift=-10pt, yshift=3pt] 
            at (img.north west) {(\thesubfigure)};
    \end{tikzpicture}
    \label{fig:circuit}
\end{subfigure}
\hfill
\begin{subfigure}{0.48\textwidth}
    \stepcounter{subfigure}
    \begin{tikzpicture}
        \node[inner sep=0pt] (img) {\includegraphics{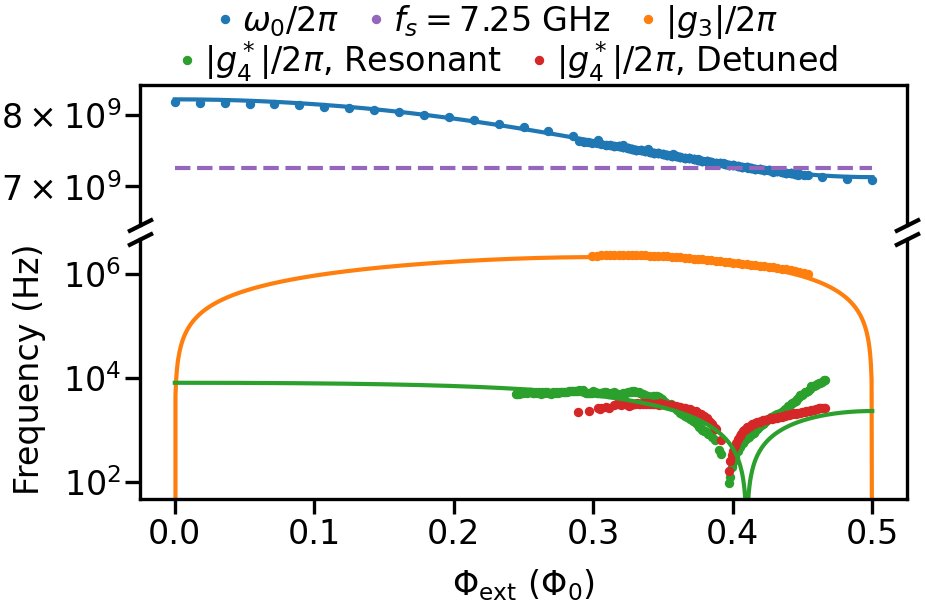}};
        \node[anchor=north west, fill=white, opacity=0.7, text opacity=1, font=\bfseries] 
            at (img.north west) {(\thesubfigure)};
    \end{tikzpicture}
    \label{fig:freqs}
\end{subfigure}
\vspace{-1.5em}
\caption{\label{fig:spa_setup} (\textsubcap{\subref{fig:circuit}}) Simplified circuit diagram of the measurement setup. During the squeezing measurement, no signal is applied to the left port of the first circulator. A 3D cavity-qubit system is connected to the output of the squeezer SPA to calibrate the output line. The output state is detected at room temperature with a heterodyne receiver.  (\textsubcap{\subref{fig:freqs}}) Measured SPA parameters (dots) compared with theory (solid lines) over the experimental flux range.}
\end{figure}

\begin{figure*}
\centering
\begin{subfigure}[t]{0.48\textwidth}
    \stepcounter{subfigure}
    \begin{tikzpicture}
        \node[inner sep=0pt] (img) {\includegraphics{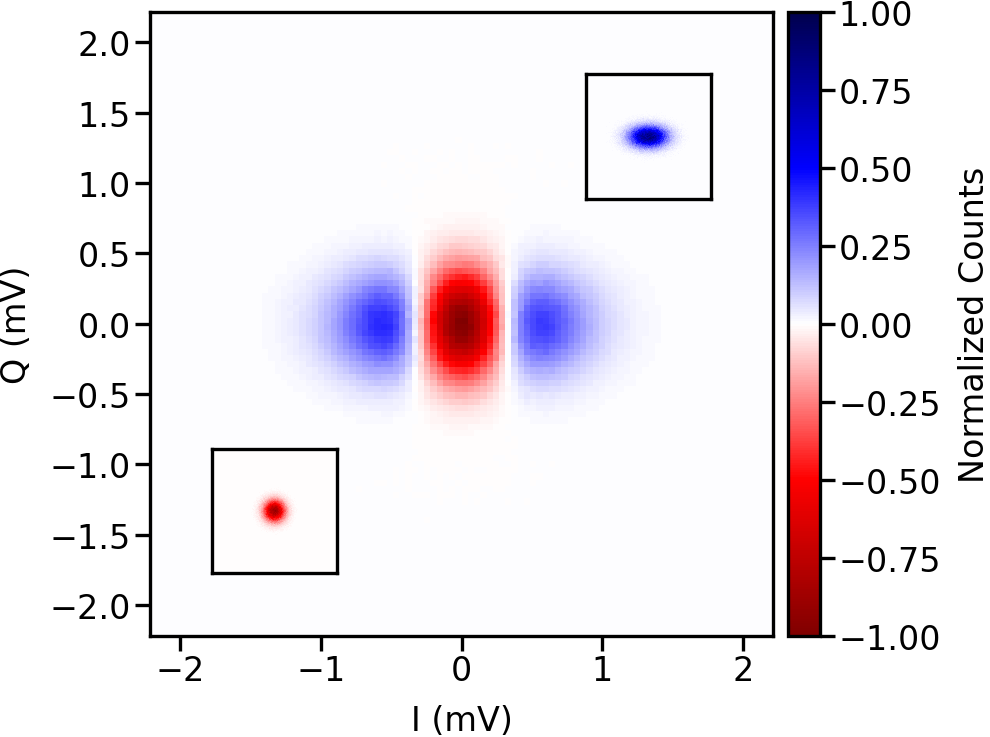}};
        \node[anchor=north west, fill=white, opacity=0.7, text opacity=1, font=\bfseries, xshift=-0pt, yshift=3pt] 
            at (img.north west) {(\thesubfigure)};
    \end{tikzpicture}
    \label{fig2:a}
\end{subfigure}
\hfill
\begin{subfigure}{0.48\textwidth}
    \vspace{0pt}
    \begin{subfigure}[t]{\linewidth}
        \stepcounter{subfigure}
        \begin{tikzpicture}
        \node[inner sep=0pt] (img) {\includegraphics{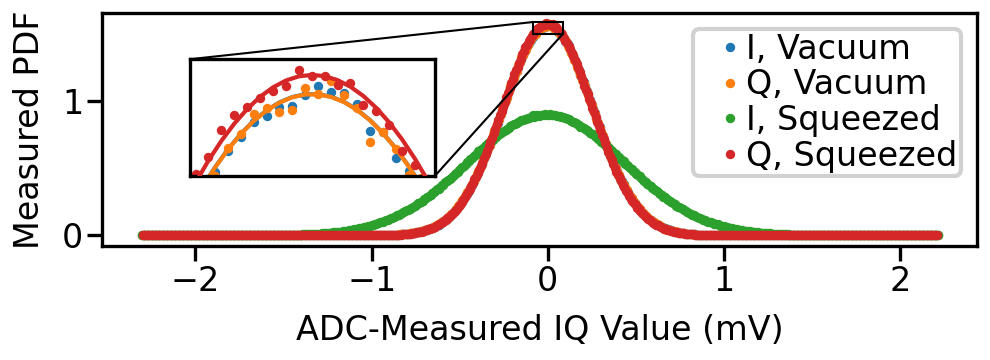}};
        \node[anchor=north west, fill=white, opacity=0, text opacity=1, font=\bfseries, xshift=25pt, yshift=-3pt] 
            at (img.north west) {(\thesubfigure)};
        \end{tikzpicture}
        \label{fig2:b}
    \end{subfigure}
    \vspace{-1.5em}
    
    \begin{subfigure}[t]{\linewidth}
        \stepcounter{subfigure}
        \begin{tikzpicture}
        \node[inner sep=0pt] (img) {\includegraphics{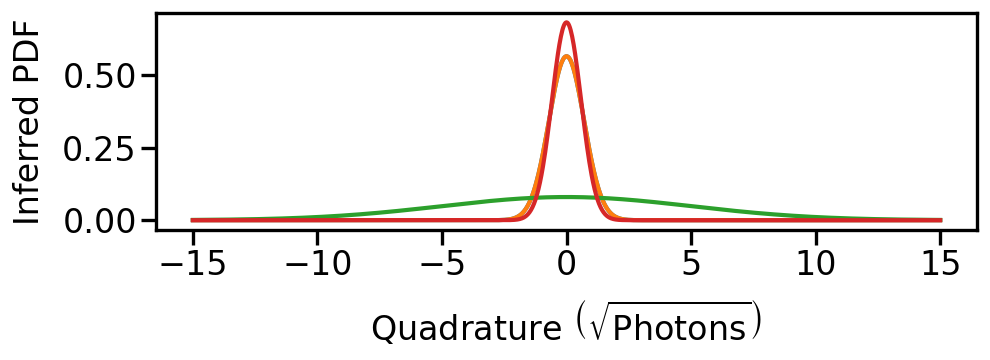}};
        \node[anchor=north west, fill=white, opacity=0, text opacity=1, font=\bfseries, xshift=35pt, yshift=-3pt] 
            at (img.north west) {(\thesubfigure)};
        \end{tikzpicture}
        \label{fig2:c}
    \end{subfigure}
\end{subfigure}

\begin{subfigure}{0.48\textwidth}
\stepcounter{subfigure}
        \begin{tikzpicture}
        \node[inner sep=0pt] (img) {\includegraphics{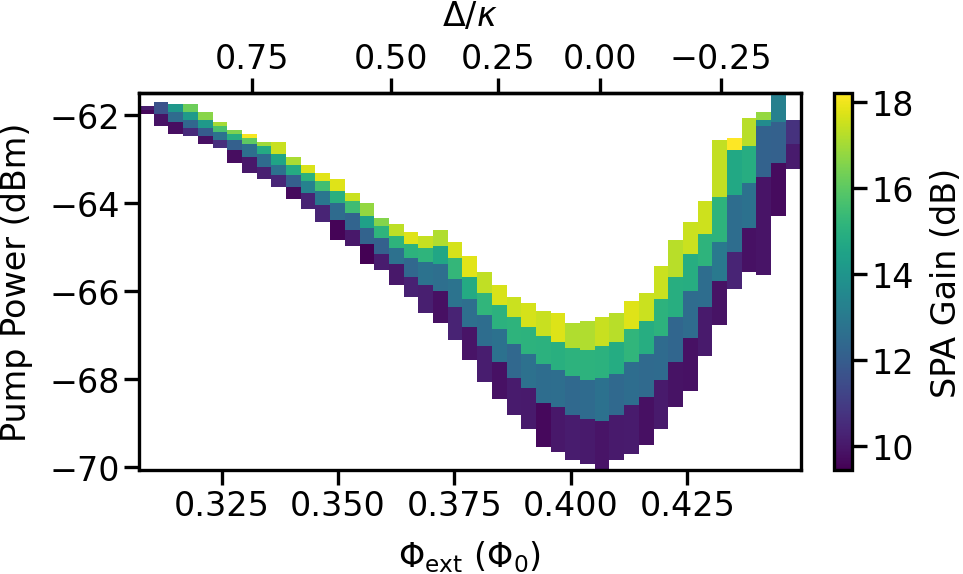}};
        \node[anchor=north west, fill=white, opacity=0.7, text opacity=1, font=\bfseries, xshift=0pt, yshift=0pt] 
            at (img.north west) {(\thesubfigure)};
        \end{tikzpicture}
        \label{fig2:d}
\end{subfigure}
\hfill
\begin{subfigure}{0.48\textwidth}
    \stepcounter{subfigure}
    \begin{tikzpicture}
        \node[inner sep=0pt] (img) {\includegraphics{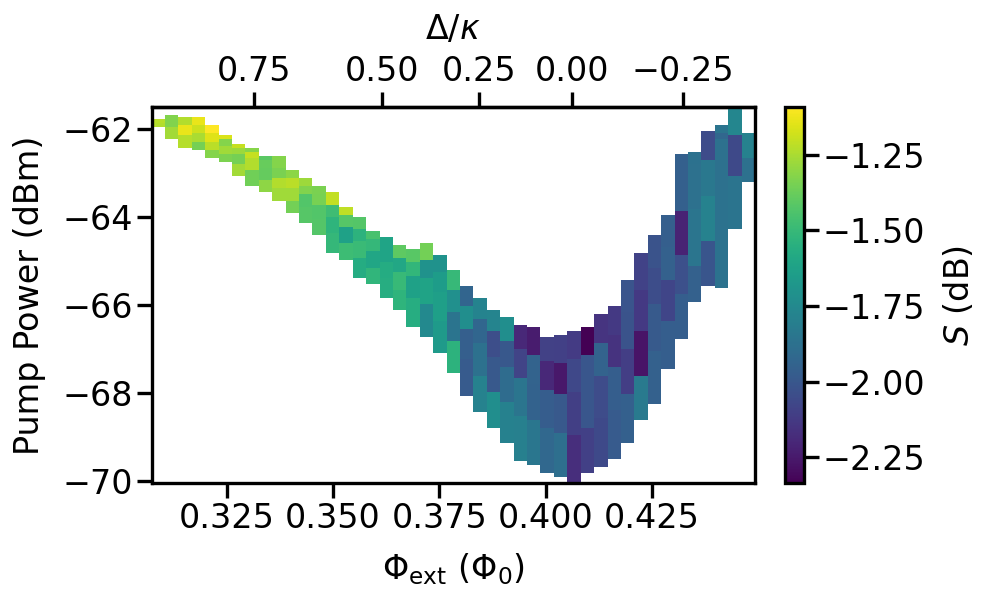}};
        \node[anchor=north west, fill=white, opacity=0.7, text opacity=1, font=\bfseries, xshift=0pt, yshift=0pt] 
            at (img.north west) {(\thesubfigure)};
    \end{tikzpicture}
    \label{fig2:e}
\end{subfigure}
\vspace{-1em}
\caption{\label{fig:main_measurement} (\textsubcap{\subref{fig2:a}}) Histogram showing the difference in counts between measured squeezed vacuum and vacuum states. A flux of counts towards large $|I|$ values and a reduction at extreme $|Q|$ values indicate phase-sensitive amplification along $I$ and squeezing along $Q$. \rev{Insets show the vacuum (bottom left) and squeezed vacuum (top right) histograms.} (\textsubcap{\subref{fig2:b}}) Line cuts of the histograms along $I$ and $Q$ with Gaussian fits. The squeezed $I$ Gaussian is wider, and squeezed $Q$ is slightly narrower, than their unsqueezed vacuum counterparts. (\textsubcap{\subref{fig2:c}}) Distributions at the qubit inferred from $\eta$, assuming Gaussian statistics. The squeezed vacuum state exhibits a $Q$ quadrature noticeably more narrow compared to the vacuum state than can be seen in (\textsubcap{\subref{fig2:b}}). (\textsubcap{\subref{fig2:d}}) Gain of the SPA as measured with a VNA. At each flux, the pump power was tuned to achieve target gains of 10, 12.5, 15, and 17.5~dB at $f_s$, defining the operating points for squeezing and IMD measurements. The pump frequency is fixed at $f_p=14.5$~GHz. (\textsubcap{\subref{fig2:e}}) Squeezing inferred from measurements at each operating point after efficiency calibration. Points with $\Phi_{\mathrm{ext}}>0.375\Phi_0$ tend to show the highest squeezing.}
\end{figure*}
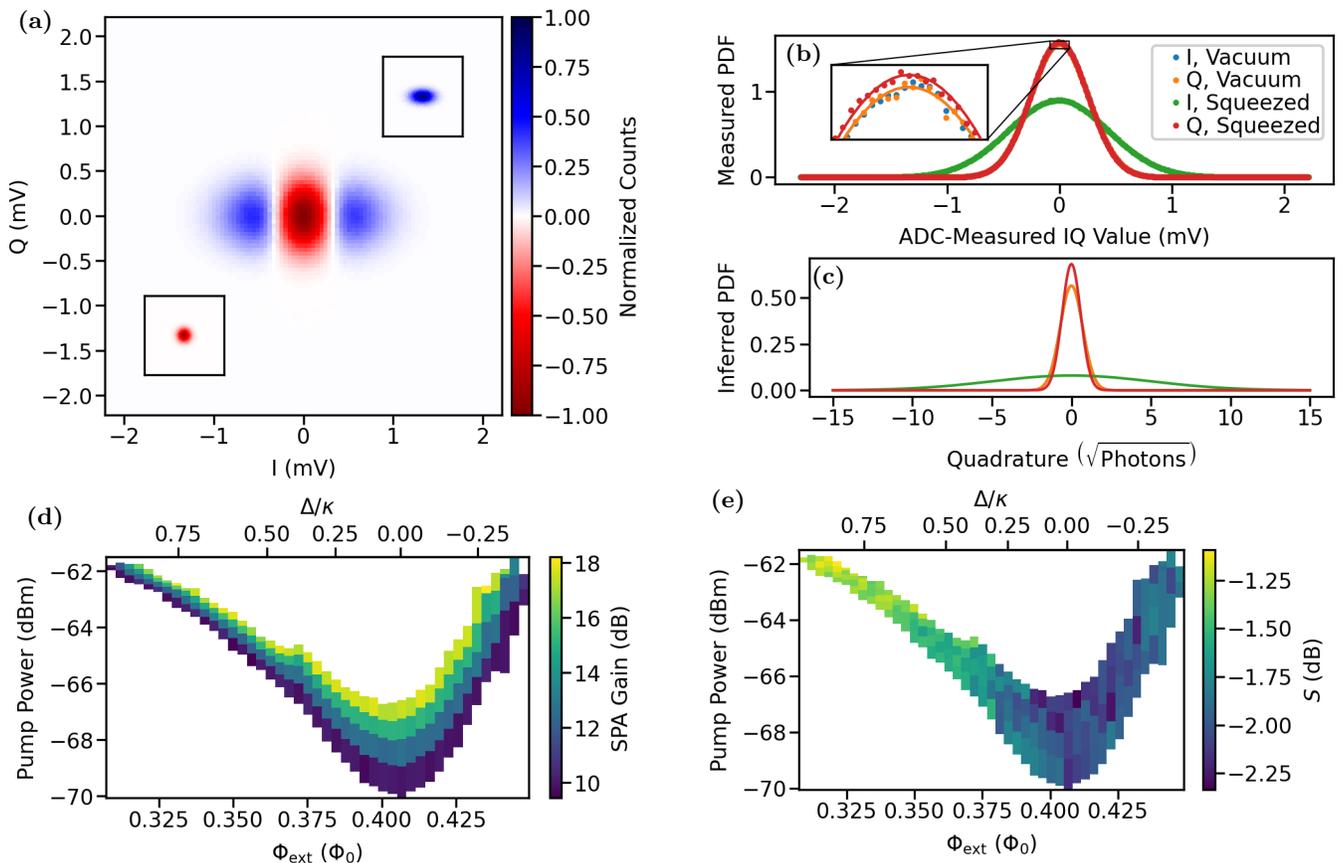

The simplified setup for the experiment is shown in Fig.~\ref{fig:spa_setup}(\subref{fig:circuit}), with the complete setup detailed in Appendix~\ref{app:exp_setup}. To calibrate the output line and define the reference plane for squeezing measurements, we use a 3D cavity–qubit system connected to the output of the SPA. The effective reference plane at which we quote the squeezing value is the coupling port of this cavity. This choice reflects the squeezing level delivered by the SPA setup to a downstream cavity, including the unavoidable loss of the circulator present in any such configuration.

\rev{Figure~\ref{fig:spa_setup}(\subref{fig:freqs}) shows the SPA parameters $\omega_0$, $|g_3|$, $|g_4^*|$ versus flux extracted from spectroscopy, parametric gain, and third-order intermodulation product ($IIP_3$) measurements, respectively, described in  Ref.~\cite{spa1}. The resonant frequency $\omega_0$ and coupling rate $\kappa$ were obtained from fits of the SPA linear response measured with a vector network analyzer (VNA).} $\kappa/2\pi$ varied between 315 and 377 MHz with flux (see Fig.~\ref{fig:kappa}). \rev{The third-order nonlinearity $|g_3|$ was calculated from the SPA gain using Eq.~6 of Ref.~\cite{spa2}, with pump frequency $f_p=14.5$ GHz and pump power $P_p$ set to yield 10~dB of gain. The fourth order nonlinearity $|g_4^*|$ was extracted from $IIP_3$ using Eqs.~\ref{eq:kerr} and \ref{iip3_body}, with the pump off, both on resonance (green) and detuned, at $f_s=7.25$~GHz (red). The $IIP_3$ measurement protocol is described further in Sec.~\ref{sec:kerr} and Appendix~\ref{app:imd}.} Overall, the SPA design parameters are similar to those used in previous readout experiments \cite{touzardGatedConditionalDisplacement2019}. Additional design details are provided in Appendix~\ref{app:spa}. As seen in Fig.~\ref{fig:spa_setup}(\subref{fig:freqs}), $|g_4^*|\approx0$ near $\omega_0=7.3$~GHz, making the neighborhood of this frequency a natural choice for studying squeezing with minimal Kerr nonlinearity. We selected a squeezing frequency $f_s = \omega_s/2\pi =  7.25$~GHz for compatibility with room-temperature electronics. 

To measure squeezing, a pump tone at frequency $f_p=2f_s$ is applied to the SPA pump port. No coherent tone is applied to the SPA input port, leaving the vacuum field incident on the resonator. The amplifier squeezes this vacuum field, and the output state is measured by a heterodyne detector at room temperature. Each measurement consisted of the following steps: (1) SPA pump turned on; (2) a 10 $\mu$s delay to allow the SPA to reach steady state; (3) a 10 $\mu$s demodulation window; (4) SPA pump turned off. Vacuum squeezing measurements (pump on) were interleaved with vacuum state measurements (pump off) using identical timing. This interleaving mitigates drift in the gain of semiconductor amplifiers~\cite{qiu_broadband_2023}. At each operating point, we acquired $10^7$ measurements of both vacuum and squeezed vacuum states.

Figure~\ref{fig:main_measurement}(\subref{fig2:a}) shows a representative histogram at one operating point. Insets display the \rev{vacuum state (bottom left) and squeezed vacuum state (top right)}, while the main panel shows their difference. Along $I$, the histogram broadens with the pump on, consistent with phase-sensitive amplification. Along $Q$, the squeezed axis in this work, the distribution narrows, indicating de-amplification of signals $\pi/2$ out of phase with the pump.
Histograms of each quadrature are shown in Fig.~\ref{fig:main_measurement}(\subref{fig2:b}). The IQ data for each state were fit to a two-dimensional normal distribution \cite{qiu_broadband_2023}, and the variances were extracted. The measured squeezing level in dB is computed as $S_{\mathrm{meas}}=10\log_{10}\frac{\sigma^2_{Q,\mathrm{on}}}{\sigma^2_{Q,\mathrm{off}}}$.

The states measured at room temperature are degraded by loss in the measurement chain and thermal noise added by the semiconductor amplifiers, particularly the HEMT amplifier \cite{qiu_broadband_2023, mallet_quantum_2011, malnouOptimalOperationJosephson2018}. The efficiency of the measurement chain $\eta$ is given by $\eta=\eta_{\mathrm{cold}}\eta_{\mathrm{hot}}$, where $\eta_{\mathrm{cold}}$ accounts for losses between squeezer and qubit-cavity, and $\eta_{\mathrm{hot}}$ accounts for losses and noise between the qubit-cavity and the room temperature detector. Internal loss of the SPA introduces an additional inefficiency, parameterized by $\eta_{\mathrm{int}} = 1-\kappa_{\mathrm{int}}/\kappa$, where $\kappa_{\mathrm{int}}$ is the internal loss rate of the SPA resonator, with $\kappa_{\mathrm{int}}/2\pi$ varying between 20 and 60 MHz with flux in our device (see Fig.~\ref{fig:kappa}). In the high-gain limit, the factor $\eta_{\mathrm{int}}$, arising from internal loss, has the same effect as external losses. Thus, the total efficiency is $\eta_{\mathrm{total}}=\eta_{\mathrm{int}}\eta_{\mathrm{cold}}\eta_{\mathrm{hot}}=\eta_{\mathrm{int}}\eta$.


If $\eta_{\mathrm{hot}}$ is known, the variances of the squeezed vacuum state at the qubit can be inferred from the measured variances (Appendix~\ref{app:loss})~\cite{qiu_broadband_2023, mallet_quantum_2011}. Moving the reference plane to the qubit provides the most accurate estimate of achievable SNR improvement for downstream applications. We measure $\eta_{\mathrm{hot}}$ using a Ramsey experiment with a detuned drive at $f_s$, yielding $\eta_{\mathrm{hot}} \approx 0.042$ (Appendix~\ref{app:qubit}). Figure~\ref{fig:main_measurement}(\subref{fig2:c}) shows the inferred distributions at the qubit plane, indicating significantly higher squeezing than that measured at room temperature.

To identify operating points with sufficient gain and suppressed Kerr, external flux $\Phi_{\mathrm{ext}}$ and pump power were swept, keeping $f_p=14.5$~GHz for $f_s=7.25$~GHz. This procedure parallels practical applications, where the squeezing frequency $f_s$ is fixed by the downstream cavity. 
At each $\Phi_{\mathrm{ext}}$, the amplifier small-signal power gain $G=\frac{|S_{11, \mathrm{on}}|^2}{|S_{11, \mathrm{off}}|^2}$ was measured with a VNA, where $S_{11, \mathrm{on}}$ and $S_{11, \mathrm{off}}$ denote the SPA voltage reflection coefficients with the pump on and off, respectively. The pump power was adjusted to achieve target gains $G$ of 10, 12.5, 15, and 17.5~dB at the fixed $f_s$. These $(\Phi_{\mathrm{ext}}, P_p)$ pairs define the operating points for subsequent measurements. Figure~\ref{fig:main_measurement}(\subref{fig2:d}) shows the gain as a function of pump power and flux,  with minimum pump power near $\Delta=0$ as expected. 
\rev{Squeezing performance at these points is shown in Fig.~\ref{fig:main_measurement}(\subref{fig2:e}). At the best operating point, we infer squeezing of $-2.3^{+0.3}_{-0.5}$~dB. Following the same procedure from room temperature measurements to qubit plane, but in $I$ rather than the $Q$ quadrature used in squeezing, we also infer anti-squeezing $S_+=17.0(4)$~dB. These two quantities can be combined to calculate the state purity $P$, which for Gaussian states is given as $P=1/\sqrt{S S_+}=0.185(50)$ with both values expressed in linear units~\cite{dassonnevilleDissipativeStabilizationSqueezing2021, qiu_broadband_2023}.}

\rev{Fig.~\ref{fig:main_measurement}(\subref{fig2:e}) reveals an asymmetry at $\Delta/\kappa = 0.2$, with significantly better squeezing where $\Delta/\kappa < 0.2$.} We next correlated the squeezing with Kerr nonlinearity measurements to determine if finite $K$ explains this asymmetry.

\bigrev{\section{Experiment Analysis}}
\bigrev{\subsection{\label{sec:kerr}Squeezing vs. Kerr}}

\begin{figure}
\centering
\begin{subfigure}{0.48\textwidth}
    \stepcounter{subfigure}
    \begin{tikzpicture}
        \node[inner sep=0pt] (img) {\includegraphics{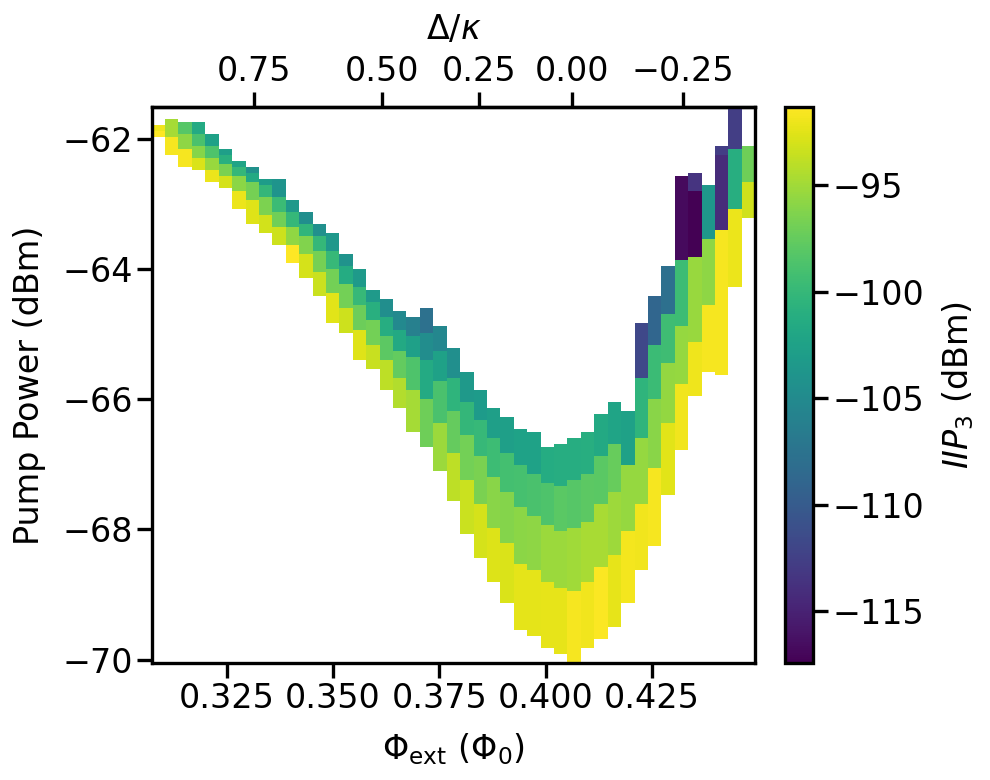}};
        \node[anchor=north west, fill=white, opacity=0.7, text opacity=1, font=\bfseries, xshift=0pt, yshift=0pt] 
            at (img.north west) {(\thesubfigure)};
    \end{tikzpicture}
    \label{fig:iip3}
\end{subfigure}

\begin{subfigure}{0.48\textwidth}
    \stepcounter{subfigure}
    \begin{tikzpicture}
        \node[inner sep=0pt] (img) {\includegraphics{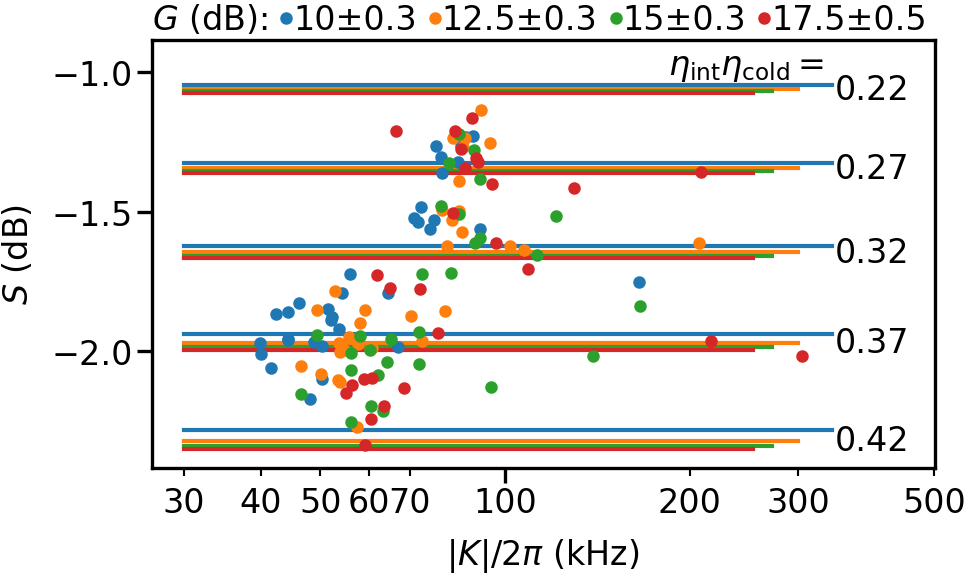}};
        \node[anchor=north west, fill=none, opacity=0.7, text opacity=1, font=\bfseries, xshift=-5pt, yshift=3pt] 
            at (img.north west) {(\thesubfigure)};
    \end{tikzpicture}
    \label{fig:s_kerr}
\end{subfigure}

\caption{\label{fig:imd_squeezing} (\textsubcap{\subref{fig:iip3}}) Input-referred third-order intercept point ($IIP_3$) measured for the SPA versus pump power and flux (bottom axis)/detuning (top axis).  (\textsubcap{\subref{fig:s_kerr}}) Comparison of squeezing and $|K|$ extracted from $IIP_3$ (circles), for several gain settings (colors: 10~dB--blue, 12.5~dB--orange, 15~dB--green, 17.5~dB--red). No clear correlation between $|K|$ and $S$ is observed. This absence of dependence is reinforced by theoretical calculations of $S$ versus $|K|$ for each gain (colored lines). Instead, the observed variation in $S$ is attributed to changes in the uncalibrated measurement efficiency, represented by clusters of colored lines corresponding to five values of $\eta_{\mathrm{int}}\eta_{\mathrm{cold}}$.}
\end{figure}

Kerr nonlinearity was characterized using an intermodulation distortion (IMD) experiment~\cite{spa1, spa2, snakeamp, biznarova_intermodulation_2024, remm_intermodulation_2023, ott_nonlinear_2004}, further described in Appendix~\ref{app:imd}. Two coherent tones of equal nominal power at frequencies $f_1$ and $f_2$, centered at $f_s + 5$~MHz and separated by 1~MHz, were applied to the SPA input. The Kerr nonlinearity drives a four-wave mixing process that generates sidebands at $2f_1 - f_2$ and $2f_2 - f_1$. The power of these sideband tones depends on input signal power and $|K|$. Sideband powers were measured with a spectrum analyzer as the input tone power was varied. The projected input power, extrapolated from low-power trends, at which the output power at $f_1$ and $f_2$ is equal to the output power at $2f_1-f_2$ and $2f_2-f_1$ defines the amplifier's input-referred third-order intercept point $IIP_3$. The value of $|K|$ can be calculated from $IIP_3$ using the equation
\rev{\begin{align}
    IIP_3&\approx \frac{\kappa_{\mathrm{ext}}^2 \hbar \omega_0}{|K|} \frac{1}{|\sqrt{G_{\mathrm{IL}}}e^{i\theta_g}+1|^3}, \label{iip3_body}
\end{align}}
derived in Appendix \ref{app:iip3}. Here $G_{\mathrm{IL}}$ is the power gain in reflection referenced to unity transmission, which differs from the common experimental definition of $G$ where the gain is referenced to the transmission with the SPA pump turned off; the ratio of $G$ to the gain referenced to unity transmission $G_{\mathrm{IL}}$ ranged between \rev{0 and 3.5~dB} at $f_s$ in this experiment. The angle $\theta_g$ is $\angle S_{11}$ at the SPA's input. At each squeezing operating point, $IIP_3$ was measured, with results shown in Fig.~\ref{fig:imd_squeezing}(\subref{fig:iip3}). 

Values of $|K|$ extracted from $IIP_3$ measurements are compared to squeezing performance in Fig.~\ref{fig:imd_squeezing}(\subref{fig:s_kerr}). For most operating points, $|K|/2\pi$ lies between 40 and 100~kHz. Outliers at higher Kerr are due to anomalous measurements which previous work \cite{snakeamp} has attributed to two-level systems (TLS). Here, it appears more likely that the anomalies are caused by higher-order amplifier nonlinearities than TLS, for reasons which are outlined in Appendix~\ref{app:imd} along with additional IMD measurement details.

The absence of operating points with substantially reduced $|K|$ is unexpected compared to \cite{spa2}, but is due to the protocol's marriage of flux to detuning, coupled with the lack of frequency resolution offered by room temperature electronics in this experiment; see Appendix~\ref{app:exp_setup}. While there are some outliers due to experimental instabilities, within the main cluster, a slight improvement in $S$ with decreasing $|K|$ is suggested.

\bigrev{\subsection{Squeezing vs. Detuning}}
\begin{figure}
\centering
\includegraphics{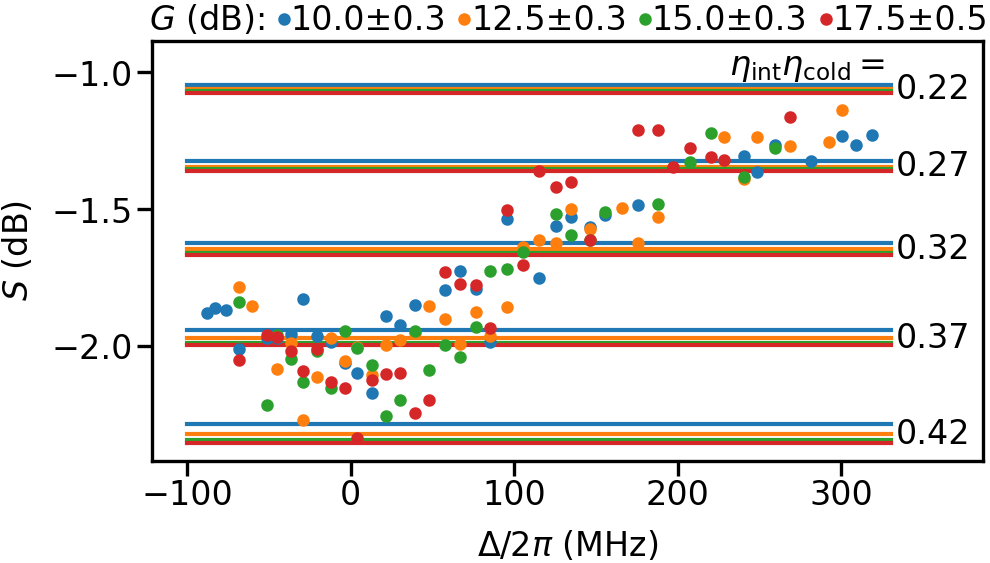}
\caption{\label{fig:delta_s} Measured squeezing versus amplifier detuning, controlled via external flux, for several gain settings (colors: 10~dB--blue, 12.5~dB--orange, 15~dB--green, 17.5~dB--red). Squeezing degrades as $|\Delta|$ increases. This degradation is attributed to changes in the uncalibrated measurement efficiency $\eta_{\mathrm{int}}\eta_{\mathrm{cold}}$ with flux. Clusters of colored lines show theoretical predictions of $S$ versus $\Delta$ for five representative values of $\eta_{\mathrm{int}}\eta_{\mathrm{cold}}$.}
\end{figure}

The experiment protocol relied on varying $\Phi_{\mathrm{ext}}$ to modify $K$ via its dependence on detuning $\Delta$. The relation between $S$ and $K$ shown in Fig.~\ref{fig:imd_squeezing}(\subref{fig:s_kerr}) is weak. However, when measured squeezing is plotted directly against $\Delta$ in Fig.~\ref{fig:delta_s}, a clear trend emerges: the best squeezing occurs near $|\Delta|=0$, with $S$ degrading monotonically as $|\Delta|$ increases. The degradation is approximately symmetric for positive and negative detuning. When $|\Delta|/2\pi$ is increased to 60~MHz, $S$ has already begun to noticeably worsen, despite the fact that this value corresponds to only $\Delta/\kappa\sim 0.2$. \rev{Crucially, this dependence is stronger than the observed dependence of $S$ on $G$ and $|K|$.}

\bigrev{\subsection{Squeezing vs. Gain}
Previous work has shown that $S$ improves with $G$ up to a point, after which further increase in gain leads to degradation and eventually $S > 0$, indicating noise above vacuum in the nominally squeezed quadrature~\cite{qiu_broadband_2023, malnouOptimalOperationJosephson2018, vaartjesStrongMicrowaveSqueezing2024}. In the present data, as seen in Fig.~\ref{fig:imd_squeezing}(\subref{fig:s_kerr}) and Fig.~\ref{fig:delta_s}, $G = 10~\mathrm{dB}$ was insufficient for optimal squeezing, while gains of 12.5, 15, and 17.5~dB resulted in similar values of $S$. A finer pump-power sweep (Appendix~\ref{app:highgain}) showed that $S$ remains nearly constant up to $G = 25~\mathrm{dB}$, after which degradation appears. The saturation up to 25~dB is consistent with the role of loss discussed in Appendix~\ref{app:spa-squeezing}, whereas the subsequent degradation is likely due to effects not captured by the present model, which assumes Gaussian fluctuations and stiff pump.}

\bigrev{\section{Comparison of Experiment and Theory}
In addition to experimental data, Fig.~\ref{fig:imd_squeezing}(\subref{fig:s_kerr}) and Fig.~\ref{fig:delta_s} show plots of predictions of squeezing from the theory of Appendix~\ref{app:spa-squeezing} for different values of $|K|$ and $\Delta$, with typical measurement parameters $g_3/2\pi=2$~MHz and $\kappa/2\pi=340$~MHz. The four target gain values of 10, 12.5, 15, and 17.5 dB are shown, as are five different values of $\eta_\mathrm{int}\eta_\mathrm{cold}$. For Fig.~\ref{fig:imd_squeezing}(\subref{fig:s_kerr}), $\Delta=0$ and $|K|$ is swept, while for Fig.~\ref{fig:delta_s}, $|K|/2\pi$ is fixed at the typical experimental value of 70~kHz and $\Delta$ is swept.}

\bigrev{The key theoretical feature is that $S$ shows no direct dependence on $|K|$ or $\Delta$. In fact, both $\Delta$ and \rev{$K$} appear only through their contributions to the Stark-shifted detuning $\Delta_{\mathrm{eff}}$, which limits the maximum achievable gain. Thus, for amplifiers tuned to the same gain $G$, the predicted squeezing performance is identical regardless of $\Delta$ or $|K|$. Only when detuning or Kerr becomes sufficiently large to prevent the amplifier from reaching the target gain does the theory predict a reduction in squeezing. This effect is shown directly in Fig.~\ref{fig:projections}(\subref{fig:projection_gain}), where the maximum achievable gain falls once $|K|$ becomes large enough that the Stark shift dominates. And in Fig.~\ref{fig:imd_squeezing}(\subref{fig:s_kerr}), the effect is represented by the theory lines growing shorter for higher gain, as theory points where the maximum gain was less than the target were not plotted.}  

\bigrev{The absence of any dependence of $S$ on $K$ in theory suggests that the variation of $S$ observed in experiment arises from other uncontrolled parameters. The most likely explanation is variation in the uncalibrated component of the measurement efficiency, $\eta_{\mathrm{int}}\eta_{\mathrm{cold}}$, across operating conditions (pump power and detuning).   Matching the observed spread in $S$ requires $\eta_{\mathrm{int}}\eta_{\mathrm{cold}}$ to vary from approximately 0.22 to 0.42 as the pump power and detuning are changed. Some mechanisms that could explain this variation are: (1) $\eta_{\mathrm{int}}=1-\kappa_{\mathrm{int}}/\kappa$ varies between $0.85$ and $0.8$ over the flux range $0.325\Phi_0$ to $0.45\Phi_0$ used to set the detuning, following the measured variation of $\kappa_{\mathrm{int}}$ (see Fig.~\ref{fig:kappa}), though by itself this variation accounts for only a small fraction of the required spread; (2) power-dependent loss~\cite{watanabePowerdependentInternalLoss2009} could increase $\kappa_{\mathrm{int}}$ with $|\Delta|$, as the pump power for a given gain tends to increase with $|\Delta|$;  (3) variation in $\eta_{\mathrm{cold}}$ may occur if flux tuning brings a lossy package mode between the SPA and qubit-cavity into resonance. All three of these mechanisms degrade squeezing through loss rather than Kerr. Capturing the spread of $S$ in the model would require treating $\kappa_\mathrm{int}$, $\kappa_\mathrm{ext}$ and $\eta_\mathrm{cold}$ as functions of the operating point rather than as constants, which we leave to future work. The identification of loss, rather than Kerr, as the factor 
limiting squeezing does not depend on which of these mechanisms dominates.}


\bigrev{While the data show no significant dependence of squeezing on $|K|$ in this setup, the sweep of operating points accessed only a factor of about two in $|K|$, and it is worth considering what a broader range of $|K|$ values would show. A measurement at a lower Kerr value would not show improved squeezing as the measured squeezing is dominated by loss in this experiment, and this would not change at lower $|K|$. At higher Kerr values, the most immediately noticeable effect would be an inability to access the desired gain setpoints due to Stark shift. This would reduce squeezing through the nominally monotonic relation between squeezing and gain. Both these regimes are depicted in
Fig.~\ref{fig:projections}(\subref{fig:projection_s}), where the predicted squeezing is independent of $|K|$ until the Stark shift prevents the amplifier from reaching the target gain. If our device could access a regime with high Kerr, with Stark shift compensated such that we could achieve sufficient gain, we could see a dependence of squeezing on Kerr due to the output state distortion illustrated by Ref.~\cite{boutinEffectHigherorderNonlinearities2017}. However, the model in Ref.~\cite{boutinEffectHigherorderNonlinearities2017} points to a ratio of $K/\kappa=10^{-2}$ limiting squeezing to approximately 5~dB. Reaching that ratio in our device, with $\kappa/2\pi \approx 340$~MHz, would require $K/2\pi\approx3$~MHz, a factor of $30$ to $75$ above the measured $40$ to $100$~kHz. While it is in principle possible to design parametric amplifiers with larger $|K|$, this runs  counter to the usual design objective of low $K/\kappa$, which is  favorable 
for high power handling. We therefore believe state-of-the-art three-wave mixing amplifiers are unlikely to enter a regime where squeezing is limited by Kerr, until losses are significantly reduced.}

\section{Conclusion}
This work has examined the performance of a SNAIL Parametric Amplifier operated as a source of single-mode vacuum squeezing under conditions representative of practical sensing and qubit-readout experiments. For a fixed squeezing frequency, squeezing performance was explored over a range of detuning and pump power near the Kerr-free point of the SNAIL, where previous experiments reported improved compression power~\cite{spa2}. Although the SPA architecture in principle allows suppression of Kerr nonlinearity through flux tuning, the measurements did not reveal operating points with significantly reduced Kerr, nor any corresponding improvement in squeezing attributable to Kerr suppression. Across the full range of operating conditions studied, the squeezing measured at a downstream system was approximately 2 dB and varied by only about $1$~dB.

Our results, together with the accompanying theoretical analysis, reinforce earlier theoretical conclusions in Ref.~\cite{boutinEffectHigherorderNonlinearities2017} that the Kerr nonlinearity $K$ alone does not ultimately determine squeezing performance. Instead, the relevant parameter is the ratio of Kerr strength to resonator linewidth, $K/\kappa$. In state‑of‑the‑art SPAs used for qubit readout, this ratio is typically on the order of $10^{-3}$, sufficiently small that squeezing is largely insensitive to the precise value of $K$. In this regime, Kerr influences squeezing only indirectly, through Stark‑shift‑induced limitations on the maximum achievable gain. In practice, gains exceeding $20$~dB are readily attainable, and Kerr therefore does not pose a significant limitation on squeezing.

\rev{In contrast to the weak dependence on Kerr, a clear and unexpected dependence of squeezing on detuning was observed for fixed gain.} Squeezing was consistently maximal near $\Delta=0$ and degraded with increasing $|\Delta|$. Theory predicts that, for devices operated at the same gain, squeezing should be independent of detuning, implying that the observed degradation arises from detuning‑dependent variations in other device parameters, such as internal and external loss. These results indicate that, under realistic operating conditions, loss dominates over Kerr in determining the achievable squeezing level for the current generation of SPA devices.

Taken together, our measurements and analysis indicate that strategies based solely on optimizing operating conditions, such as pump power and detuning, are unlikely to yield substantially improved squeezing in practical SPAs, particularly when the squeezing frequency is fixed by a downstream device. Instead, achieving higher squeezing levels will require improvements in internal resonator quality factors and reductions in package‑ and wiring‑related loss. We expect these conclusions to apply broadly to flux‑tunable, three‑wave‑mixing parametric amplifiers and to provide useful guidance for the design of future experiments employing squeezed microwave states in superconducting quantum systems.  
\section{Acknowledgments}
We acknowledge, for valuable discussions and hardware support, Ryan Kaufman, Maria Nowicki, Joe Aumentado, Florent Lecocq, Om Joshi, and Xu Han. This work was supported by the Army Research Office (Grant No. W911NF-23-1-0251 and W911NF-23-1-0096), the Air Force Office of Scientific Research (Grant No. FA9550-22-1-0203)  and the Defense Advanced Research Projects Agency (Grant No. HR00112420343). A.K. and D. M. were supported by Department of Energy under grant number DESC0019461 and by National Science Foundation under grant number DMR-2508447. H.M.C was supported in part by an appointment to the Department of Defense (DOD) Research Participation Program administered by the Oak Ridge Institute for Science and Education (ORISE) through an interagency agreement between the U.S. Department of Energy (DOE) and the DOD. ORISE is managed by Oak Ridge Associated Universities (ORAU) under DOE contract number DE-SC0014664. Sample fabrication was performed in the University of Texas at Austin Microelectronics Research Center, a member of the National Nanotechnology Coordinated Infrastructure (NNCI), which is supported by the National Science Foundation (Grant No. ECCS-2025227). All opinions expressed in this paper are the author's and do not necessarily reflect the policies and views of DOD, DOE, or ORAU/ORISE.

\rev{\section{Data Availability}}
\rev{The data that support the findings of this article are openly available~\cite{zenodo}. }

\appendix

\section{SQUEEZING THEORY IN THE DEGENERATE REGIME} \label{app:spa-squeezing}
\input{squeezingTheoryAppendix}

\section{FULL EXPERIMENTAL SETUP} \label{app:exp_setup}

\begin{figure*}[!htbp]
\centering
\includegraphics{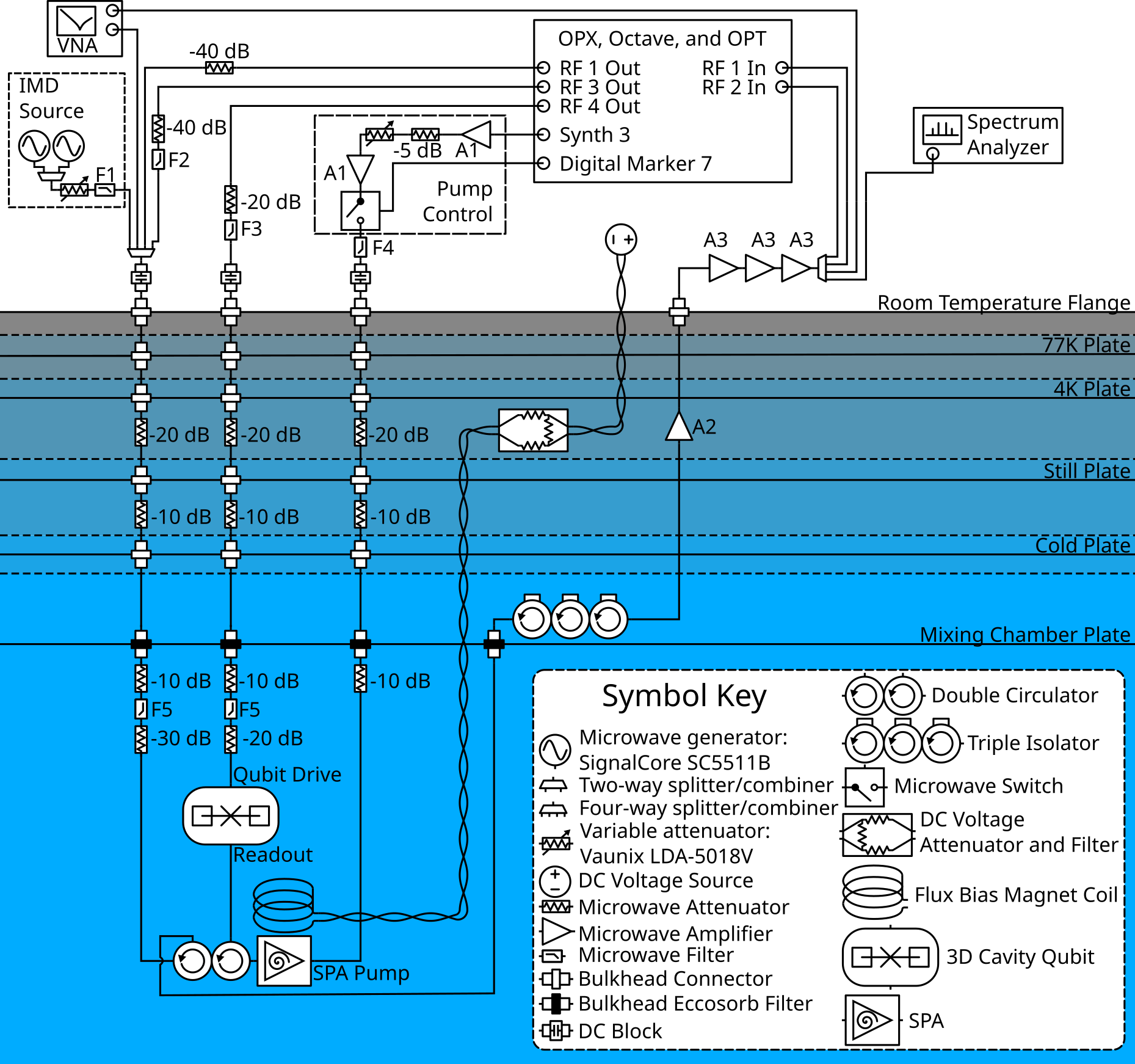}
\label{fig:full_circuit}

\caption{\label{fig:full_hw} Hardware setup used in this experiment. The Quantum Machines OPX, Octave, and OPT are grouped into a single symbol. Inset is the symbol key for the experimental setup. The VNA is a Copper Mountain Technologies M5180, and the Spectrum Analyzer is a Keysight N9000B; a Stanford Research Systems FS725/3 provides the reference clock signal for the experiment.}
\end{figure*}


The complete hardware configuration is shown in Fig.~\ref{fig:full_hw}, with the part numbers of filters and amplifiers provided in Table~\ref{tab:bom}. In initial tests, referencing the SPA pump and demodulation tone generators to a 10~MHz clock signal did not yield the phase stability required for accurate squeezing measurements. To overcome this limitation, a Quantum Machines Octave was employed in a nonstandard configuration. One of the Octave's internal microwave generators, ordinarily used as the local oscillator for a mixer, was repurposed and used directly as the pump source. The pump power was adjusted using a variable attenuator in combination with a discrete microwave switch.

This approach ensured that both the pump and demodulation tones were referenced to the Octave internal 1~GHz clock, providing the required phase stability over long measurement intervals. However, this configuration imposes a constraint: the signal frequency $f_s$ and the intermediate frequency $f_{\mathrm{IF}}$ must be integer multiples of 250~MHz. If this condition was not satisfied, the resulting demodulated signal exhibited phase drift and jitter across the ensemble of measurements at a given operating point, degrading the extracted squeezing value.

\begin{table}[!tbhp]
    \centering
    \begin{tabular}{|l|l|l|}
        \hline
        \textbf{Label} & \textbf{Manufacturer} & \textbf{Model Number} \\
        \hline
        A1 & Quantic X-Microwave & XM-A342-0804C-1 \\
        A2 & Low Noise Factory & LNF-LNC4\_16B \\
        A3 & Mini-Circuits & ZX60-123LN-S+ \\
        F1 & Mini-Circuits & ZBSS-9G-S+ \\
        F2 & Quantic X-Microwave & XM-C714-0404C \\
        F3 & Mini-Circuits & ZBSS-6G-S+ \\
        F4 & Mini-Circuits & ZHSS-11G-S+ \\
        F5 & K\&L Microwave & 6L250-12000/TP2600-OP/O \\
        \hline
    \end{tabular}
    \caption{\label{tab:bom}Manufacturers and part numbers for filters and amplifiers used in the experiment.}
\end{table}

\section{SPA DESIGN}\label{app:spa}
The SPA used in this experiment is an evolution from  device C of Ref.~\cite{spa1}. To increase the junction participation ratio $p$, we reduced the resonator length to 1 mm and increased its width to 400 $\mu$m, giving a resonator impedance $Z_c=50$~$\Omega$. Fabrication variance resulted in the unexpectedly low $\alpha=0.05$; with $L_J=44$~pH, the SPA has $p=0.7$. The outcome, relative to device C of Ref.~\cite{spa1}, is an SPA with slightly higher maximum resonance frequency of 8.2~GHz and a similar tunable frequency range of 1.1~GHz.

The SPA was fabricated on silicon with a thickness of 500 $\mu$m. The junctions are fabricated with a Dolan bridge process \cite{dolan_offset_1977}, with a 35~nm aluminum base layer, an  oxidation step at 6~Torr for 5~minutes, a 120~nm Al top layer, followed by a cap oxidation step at 10 torr for 5 minutes. The chip had a normal metal ground plane on its backside realized by depositing a Ti sticking layer followed by 2 $\mu$m silver. The chip ground plane is bonded to the copper ground plane of the printed circuit board with silver paint. The circuit board was then soldered to a gold-plated aluminum housing and enclosed in a mu-metal shield. Flux is applied to the SPA with a superconducting coil wound on a copper spool positioned below the chip. 

To obtain the theoretical curves in Fig.~\ref{fig:spa_setup}(\subref{fig:freqs}), the resonant frequency of the SPA was fit to equation (F22) of Ref.~\cite{spa2}, with line inductance and capacitance calculated from the resonator geometry, and asymmetry parameter $\alpha$ and junction inductance $L_J$ left as free parameters in the fit. The best-fit $\alpha$ and $L_J$ values were then used to calculate $g_3$ and $g_4^*$ according to equations (F36) and (F34) of Ref.~\cite{spa2}. 

\begin{figure}[b]
\centering
\includegraphics{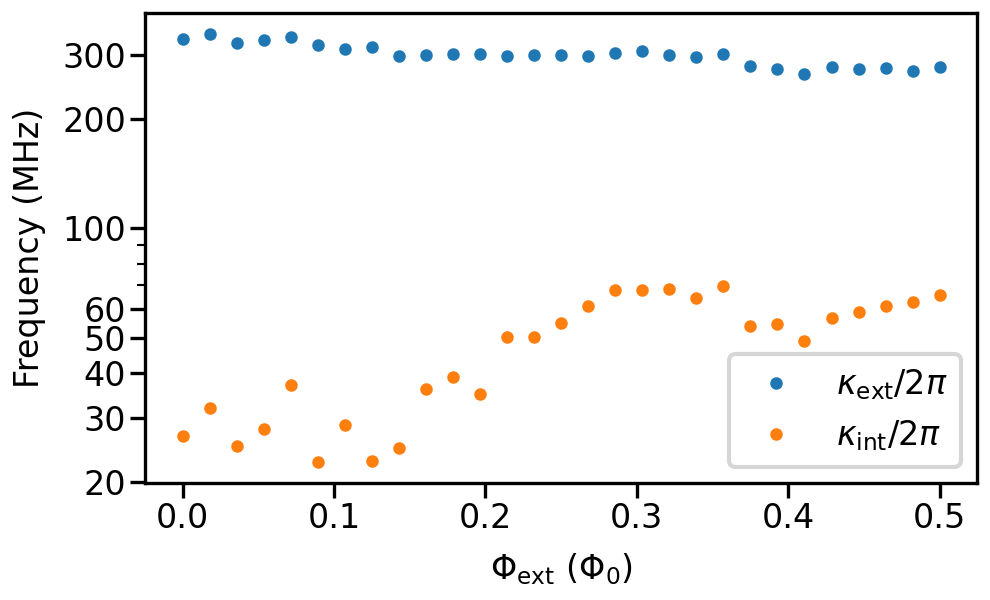}
\label{fig:iip3}

\caption{\label{fig:kappa} External $\kappa_\mathrm{ext}$ and internal $\kappa_\mathrm{int}$ coupling rates for the SPA used in this experiment.}
\end{figure}

Figure~\ref{fig:kappa} shows the resonator external $\kappa_\mathrm{ext}$ and internal $\kappa_\mathrm{int}$ coupling rates extracted from fitting the SPA signal port voltage reflection coefficient $S_{11}$, measured with a VNA, to a parallel RLC resonator model. With $\kappa = \kappa_\mathrm{int} +\kappa_\mathrm{ext}$, we find that $\kappa_{\mathrm{int}}/\kappa$ is in the range of 0.1 to 0.2, and thus $\eta_\mathrm{int} = 1-\kappa_\mathrm{int}/\kappa$ ranges from 0.8 to 0.9. 

\section{EFFECT OF LOSS}\label{app:loss}

\begin{figure}[!htbp]
\centering
\begin{subfigure}{0.48\textwidth}
    \stepcounter{subfigure}
    \begin{tikzpicture}
        \node[inner sep=0pt] (img) {\includegraphics{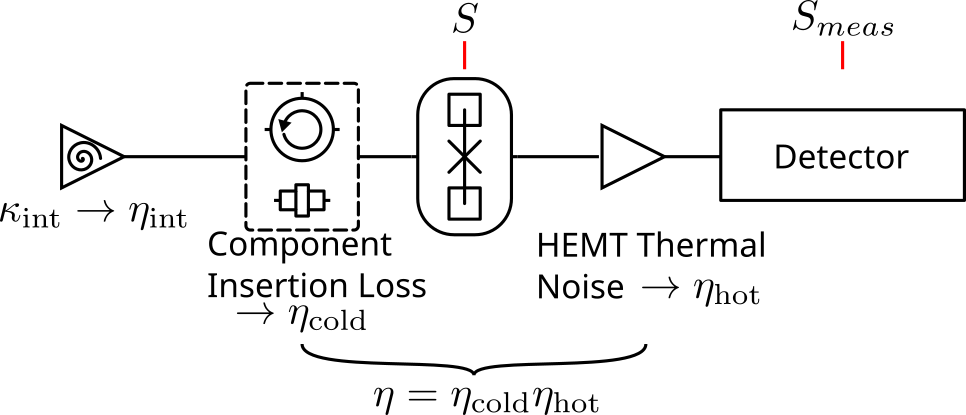}};
        \node[anchor=north west, fill=none, opacity=0.7, text opacity=1, font=\bfseries, xshift=-5pt, yshift=3pt] 
            at (img.north west) {(\thesubfigure)};
    \end{tikzpicture}
    \label{fig:setup_cartoon}
\end{subfigure}

\begin{subfigure}{0.48\textwidth}
    \stepcounter{subfigure}
    \begin{tikzpicture}
        \node[inner sep=0pt] (img) {\includegraphics{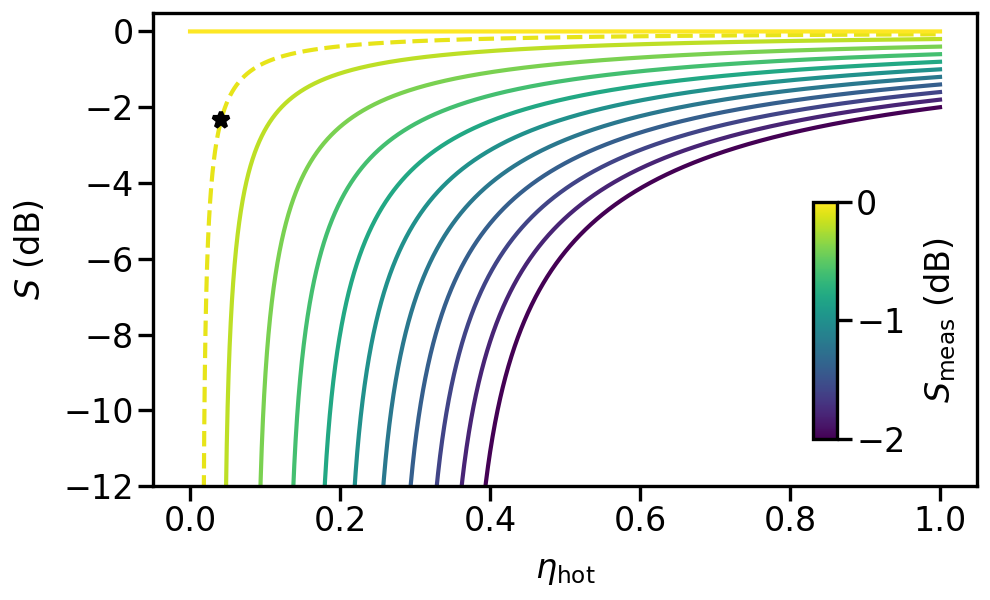}};
        \node[anchor=north west, fill=white, opacity=0.7, text opacity=1, font=\bfseries, xshift=0pt, yshift=0pt] 
            at (img.north west) {(\thesubfigure)};
    \end{tikzpicture}
    \label{fig:loss}
\end{subfigure}

\caption{\label{fig:loss_main} (\textsubcap{\subref{fig:setup_cartoon}}) Simplified diagram of experiment setup, with primary contributors to inefficiency at each stage shown, along with squeezing reference planes. (\textsubcap{\subref{fig:loss}}) Mapping from $S_{\mathrm{meas}}$ to $S$, the squeezing at the qubit, based on $\eta_{\mathrm{hot}}$. $\eta_{\mathrm{hot}}$ and the best $S_{\mathrm{meas}}$ for this experiment are indicated by a star. }
\end{figure}

The level of vacuum squeezing measured at room temperature, $S_{\mathrm{meas}}$, is expected to be significantly worse than the squeezing at the SPA output due to loss in microwave components and noise added in semiconductor amplifiers, particularly the HEMT amplifier in the first amplification stage. A simplified diagram of the experimental setup showing the measurement efficiencies of various stages and the primary contributors of inefficiency is shown in Fig.~\ref{fig:loss_main}(\subref{fig:setup_cartoon}). The measurement efficiency of the output line $\eta_{\mathrm{hot}}$ can be estimated as described in Appendix~\ref{app:qubit}, which can then be used to extract $S$ from $S_{\mathrm{meas}}$.

Following the approach of Mallet \textit{et al.} and Qiu \textit{et al.} \cite{qiu_broadband_2023, mallet_quantum_2011}, \begin{equation}
    S=10\log_{10}\left(\frac{10^{S_{\mathrm{meas}}/10}-(1-\eta_{\mathrm{hot}})}{\eta_{\mathrm{hot}}}\right)
\end{equation}
with both $S$ and $S_{\mathrm{meas}}$ expressed in dB. If $\eta=\eta_{\mathrm{cold}}\eta_{\mathrm{hot}}$ were measured, the same procedure would yield the squeezing at the SPA, rather than at the qubit.

In this experiment, the value of $\sigma_{Q, \mathrm{on}}^2/\sigma^2_{Q,\mathrm{off}}$ typically settled to within 0.05\% of its final value within $6.5\times 10^6$ of the sampled $10^7$ shots. Given the $\eta_{\mathrm{hot}}$ and best $S_{\mathrm{meas}}$ measured in the main experiment, a conservative assumption of 10\% uncertainty in estimating $\eta_{\mathrm{hot}}$ and 0.05\% in $S_{\mathrm{meas}}$ would lead to \rev{$-2.0$ dB $\geq S \geq -2.8$ dB}. Almost all of the uncertainty arises from the $\eta_{\mathrm{hot}}$ estimation. Since the same value for $\eta_{\mathrm{hot}}$ is used in calculation of all of the squeezing values in the experiment, there could be a systematic underestimation or overestimation of $S$, but not large point-to-point swings.

\section{QUBIT MEASUREMENTS}\label{app:qubit}
\begin{figure}
\centering

\begin{subfigure}{0.48\textwidth}
    \stepcounter{subfigure}
    \begin{tikzpicture}
        \node[inner sep=0pt] (img) {\includegraphics{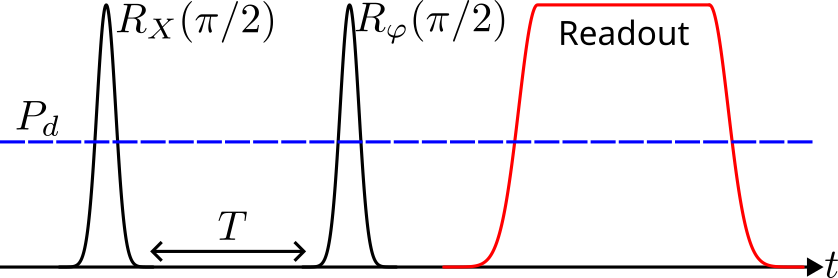}};
        \node[anchor=north west, fill=white, opacity=0.7, text opacity=1, font=\bfseries, xshift=0pt, yshift=0pt] 
            at (img.north west) {(\thesubfigure)};
    \end{tikzpicture}
    \label{fig:ramsey_pulses}
\end{subfigure}

\begin{subfigure}{0.48\textwidth}
    \stepcounter{subfigure}
    \begin{tikzpicture}
        \node[inner sep=0pt] (img) {\includegraphics{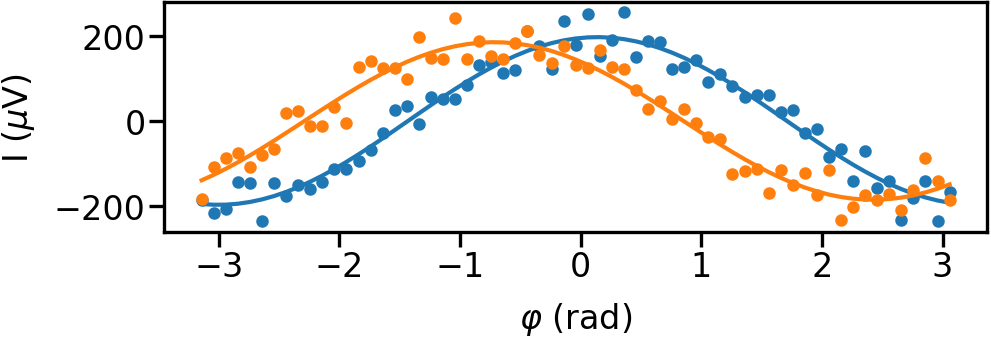}};
        \node[anchor=north west, fill=white, opacity=0.7, text opacity=1, font=\bfseries, xshift=0pt, yshift=0pt] 
            at (img.north west) {(\thesubfigure)};
    \end{tikzpicture}
    \label{fig:ramsey_sines}
\end{subfigure}

\begin{subfigure}[t]{0.2\textwidth}
    \stepcounter{subfigure}
    \begin{tikzpicture}
        \node[inner sep=0pt] (img) {\includegraphics{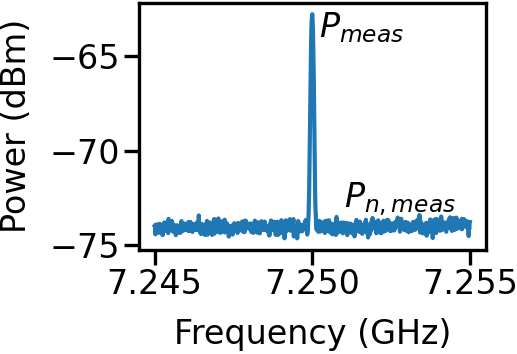}};
        \node[anchor=north west, fill=white, opacity=0.7, text opacity=1, font=\bfseries, xshift=35pt, yshift=-3pt] 
            at (img.north west) {(\thesubfigure)};
    \end{tikzpicture}
    \label{fig:specan}
\end{subfigure}
\hspace{17pt}
\begin{subfigure}[t]{0.2\textwidth}
    \stepcounter{subfigure}
    \begin{tikzpicture}
        \node[inner sep=0pt] (img) {\includegraphics{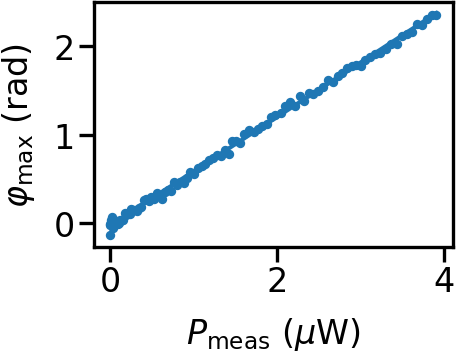}};
        \node[anchor=north west, fill=white, opacity=0.7, text opacity=1, font=\bfseries, xshift=25pt, yshift=-3pt] 
            at (img.north west) {(\thesubfigure)};
    \end{tikzpicture}
    \label{fig:ramsey_fit}
\end{subfigure}
\vspace{-1.5em}
\caption{\label{fig:ramsey} (\textsubcap{\subref{fig:ramsey_pulses}}) Pulse sequence for the Stark-shifted Ramsey measurement. (\textsubcap{\subref{fig:ramsey_sines}}) Representative Ramsey measurements for two different detuning pulse generator powers, shown as different colors, with a solid line depicting the fit for each generator power. For each generator power, the change in $\varphi$, relative to the $P_{d}=0$ case, that yields the maximum $I$ is stored as $\varphi_{\mathrm{max}}$. (\textsubcap{\subref{fig:specan}}) A spectrum analyzer measurement. The power of the peak at $f_{d}$ is $P_{\mathrm{meas}}$, while $P_{\mathrm{n, meas}}$ is the noise power measured away from the peak. (\textsubcap{\subref{fig:ramsey_fit}}) Stark-shifted Ramsey data (circles) showing good conformance to a linear fit (solid line). The y-intercept of the fit is used to determine $\theta_{max}(P_d=0)$, and the slope informs $G_{line}$.}
\end{figure}

Present on the output of the SPA was a 3D cavity-qubit system with readout resonator coupling rate $\kappa_r/2\pi=2.12$ MHz, cavity-qubit coupling rate $\chi/2\pi=1.88$ MHz, and relaxation time $T_1=18$ $\mu$s and Ramsey coherence time $T_{2R}=22$ $\mu$s. To measure $\eta_{\mathrm{hot}}$, Ramsey measurements with variable phase $\varphi$ on the second $\pi/2$ pulse were taken with a $\omega_d=\omega_s = 2\pi \cdot 7.25$ GHz tone applied to the readout port; see Fig.~\ref{fig:ramsey}. This tone populated the readout resonator with some number of photons $n_d$, which detuned the qubit proportional to $\chi$. This shift in qubit frequency manifested as a shift $\Delta_{\varphi_{\mathrm{max}}}={n}_d \chi T$ in the angle $\varphi_{\mathrm{max}}$ that returned the greatest probability of measuring the qubit in $\ket{0}$ after the second $\pi/2$ pulse, relative to $\varphi_{\mathrm{max}}(P_d=0)$, with a wait time $T=10$~$\mu$s between $\pi/2$ pulses. The power incident on the readout port was then calculated as 
\begin{equation}
    P_{d}=\frac{\hbar \omega_d \Delta_{\theta_{\mathrm{max}}} (\omega_{c,e}+\omega_{c,g} - 2\omega_d)^2}{4 T \chi \kappa_r},
\end{equation}
$\omega_{c,e}$ $(\omega_{c,g})$ being the readout cavity frequency with the qubit in the excited (ground) state.

Simultaneously, the noise spectrum around $\omega_d$ was measured with a spectrum analyzer. The measured power at $\omega_d$ allowed determination of the total output line gain $G_{\mathrm{line}}=P_{\mathrm{meas}}/P_d$ from qubit readout port to spectrum analyzer. The noise power $P_{\mathrm{n,meas}}$, was averaged over a band from $\omega_d/2\pi\pm1$~MHz to $\omega_d/2\pi\pm50$~MHz. The system noise temperature was then calculated as $T_{\mathrm{sys}}=P_{\mathrm{n, meas}}/k_\mathrm{B} G_{\mathrm{line}} B$, $B$ being the resolution bandwidth of the spectrum analyzer; here, the measured noise power was normalized to $B=1$ Hz by the Noise Marker function of the spectrum analyzer. The system noise temperature is related to $\eta_{\mathrm{hot}}$ by \cite{mallet_quantum_2011} 
\begin{equation}
    \eta_{\mathrm{hot}}=\frac{1}{1+2k_B T_{\mathrm{sys}}/h f_s} .\label{eq:eta_hot}
\end{equation}
We note that Ref.~\cite{qiu_broadband_2023} defined $\eta_{\mathrm{hot}} = h f_s/2k_B T_{\mathrm{sys}}$, which is the asymptote of Eq.~(\ref{eq:eta_hot}) in the limit $k_B T_{\mathrm{sys}} >> h f_s$, which applies when the first amplifier in the measurement chain is a HEMT amplifier.

Finally, the qubit was also used to verify that the state incident on the SPA was the vacuum state rather than a thermal state. $\frac{1}{T_{2R}}=\frac{1}{2T_1}+\frac{1}{T_\phi}$, with $\frac{1}{T_\phi}$ being the total dephasing rate due to thermal noise and all other mechanisms. That is, $\frac{1}{T_\phi}\geq \Gamma_\phi^{th}$ with $\Gamma_{\phi}^{th}=\frac{\bar{n}_{\mathrm{th}}\kappa \chi^2}{\kappa^2 + \chi^2}$. Our measurements gave a value $\bar{n}_{th} \leq 0.018 \ll 0.5$, giving important confirmation of a vacuum state exiting the SPA with the pump off. If the SPA state with the pump off had been thermal rather than a vacuum state, the measured squeezing would appear more significant, as inefficiency in the measurement chain has a lessened effect on the larger squeezed thermal state. But while the measured squeezing would appear better, the squeezed thermal state would not necessarily offer an SNR improvement over a vacuum state in a downstream experiment.

\section{IMD MEASUREMENT}\label{app:imd}

\begin{figure}

\centering
\begin{subfigure}{0.48\textwidth}
    \stepcounter{subfigure}
    \begin{tikzpicture}
        \node[inner sep=0pt] (img) {\includegraphics{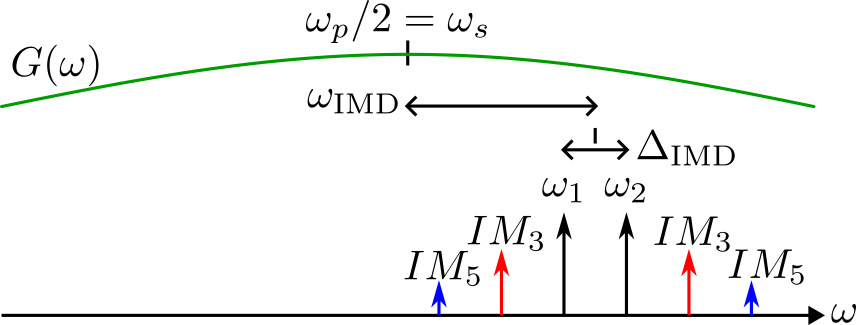}};
        \node[anchor=north west, fill=white, opacity=0, text opacity=1, font=\bfseries, xshift=0pt, yshift=5pt] 
            at (img.north west) {(\thesubfigure)};
    \end{tikzpicture}
    \label{fig:imd_pulses}
\end{subfigure}

\begin{subfigure}{0.48\textwidth}
    \stepcounter{subfigure}
    \begin{tikzpicture}
        \node[inner sep=0pt] (img) {\includegraphics{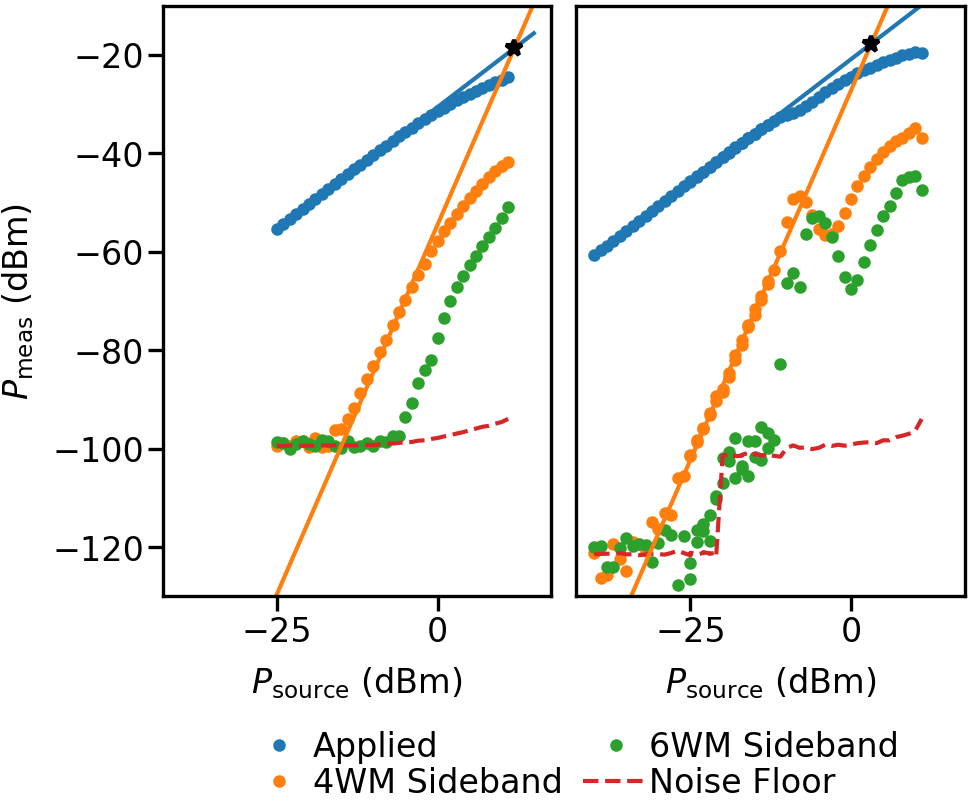}};
        \node[anchor=north west, fill=white, opacity=0, text opacity=1, font=\bfseries, xshift=38pt, yshift=0pt] 
            at (img.north west) {(\thesubfigure)};
        \stepcounter{subfigure}
        \node[anchor=north, fill=white, opacity=0, text opacity=1, font=\bfseries, xshift=30pt, yshift=0pt] 
            at (img.north) {(\thesubfigure)};
        
    \end{tikzpicture}
    \label{fig:imd_clean}
\end{subfigure}

\caption{\label{fig:imd} (\textsubcap{\subref{fig:imd_pulses}}) IMD measurement scheme. The applied tones in black interact with the amplifier's Kerr nonlinearity and produce the $IM_3$ tones, as well as higher order processes producing $IM_5$ and higher tones. (\textsubcap{b}) A well-behaved IMD measurement. (\textsubcap{c}) A typical IMD measurement with unexpected feature, from a different amplifier operating point. The jump in the noise floor at $P_\textrm{source} = -20$~dBm is due to the change of resolution bandwidth during the sweep in order to measure lower powers precisely.}

\end{figure}
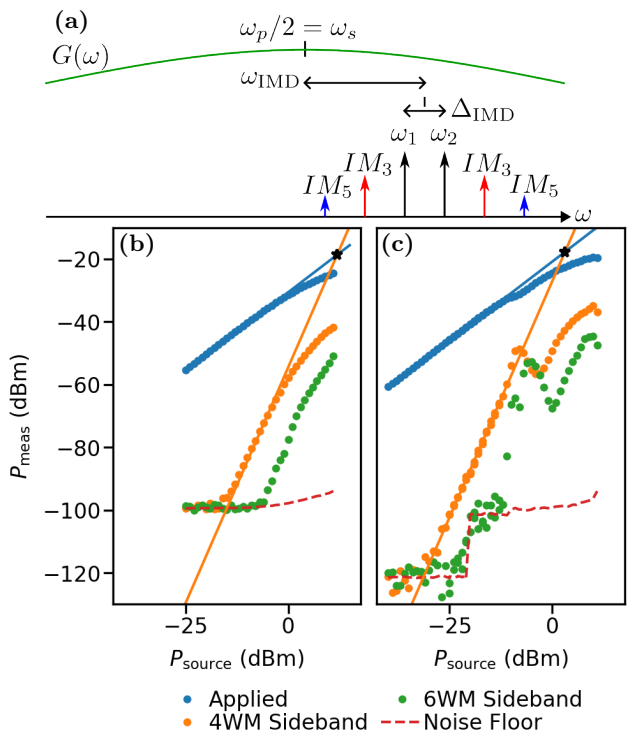

The 4-wave-mixing nonlinearity $K$ of the amplifier was characterized by an inter-modulation distortion experiment. See Fig.~\ref{fig:imd} for details. In this measurement, two microwave tones with the same power are applied to the SPA's input. The input tones have frequencies $\omega_1$ and $\omega_2$ with a frequency separation $\Delta_{\mathrm{IMD}}/2\pi=1$~MHz and a center frequency detuned from the gain center frequency by $\omega_{\mathrm{IMD}}=\omega_p/2 - (\omega_1+\omega_2)/2=2\pi\cdot5$~MHz. $K$ causes the tones to interact and produce $IM_3$ sideband tones at $\omega_1 - \Delta_{\mathrm{IMD}}$ and $\omega_2 + \Delta_{\mathrm{IMD}}$. By measuring the relative strengths of the output tones at $\omega_1$, $\omega_2$, $\omega_1-\Delta_{\mathrm{IMD}}$ and $\omega_2+\Delta_{\mathrm{IMD}}$, $K$ can be determined. 

A typical IMD measurement with the amplifier off has the measured power of the $\omega_1$ and $\omega_2$ tones increasing by 1 dB, and $IM_3$ tone power increasing by 3 dB, as the source power increases by 1 dB. Eventually, the amplifier saturates and the rate of increase declines. Such behavior is shown in Fig.~\ref{fig:imd}b. The power at which the 1 dB/dB and 3 dB/dB lines intersect, shown as the star in the figure, is known as the third-order intercept point, and when the power at the input of the amplifier is determined from this intercept point, the result is the input-referred third-order intercept point, or $IIP_3$.

With the amplifier on, the IMD behavior becomes significantly stranger. Generally speaking, there is a region at low source power where the tones increase by 1 dB and 3 dB per 1 dB of source power. But as the source power increases, there are frequently unexpected declines and increases in the measured powers as seen in Fig.~\ref{fig:imd}(c). Similar behavior has been observed in another superconducting parametric amplifier, and was explained as being due to two-level systems (TLS) providing another source of fourth-order nonlinearity which dominates at lower powers. Amplifier $K$ is explained as being lower in magnitude with an opposite sign, and dominating at higher powers once TLS nonlinearity has saturated \cite{snakeamp}. Importantly, the TLS provide a fourth-order nonlinearity by coupling to low-frequency beat patterns in the sum of the two applied tones. This implies that the TLS will contribute confounding nonlinearity to IMD measurements, but not to a squeezing measurement, which does not employ two tones relatively close in frequency. As a squeezing measurement will be comparable in power to the low end of practically measurable IMD powers, it is important to determine whether IMD at low powers is dominated by TLS or $K$ in our case.

Through observation of these features, we have arrived at the conclusion that the IMD artifacts in our experiment are unlikely to be due to TLS for three reasons. First, the artifacts are broadly insensitive to $\Delta_{IMD}$; in the case of the ``snake'' amplifier \cite{snakeamp}, the artifacts were noticeable at $\Delta_{\mathrm{IMD}}/2\pi\leq 100$ kHz, but disappeared for $\Delta_{\mathrm{IMD}}/2\pi\geq 1$ MHz. In our experiment, increasing $\Delta_{\mathrm{IMD}}/2\pi$ to 1 MHz had little effect, and even at $\Delta_{\mathrm{IMD}}/2\pi=12.66$ MHz, a measurement that required applying one tone and allowing the signal to mix with the idler, the artifact was still prominent. This would imply a TLS $T_2\ll 80$ ns if a TLS was the source of this artifact.

Second, the artifacts seen in this experiment do not have a shape that fits the proposed TLS model. It was common to see artifacts where the sideband power would increase by more than 3 dB/dB immediately before a dip; the TLS model explains dips by a cancellation of fourth-order nonlinearity due to the TLS nonlinearity having an opposite sign from $K$, but this cannot explain an increase in sideband power before a dip, as any fourth-order nonlinearity should only be capable of increasing sideband power by \rev{3~dB/dB} regardless of its origin. In addition, at some operating points two dips in the sideband power can be seen. If this is due to the interaction of multiple TLS, it would require that at least two TLS provide nonlinearity contributions with opposite signs, allowing a cancellation as one TLS saturates, and another cancellation as the second TLS saturates. We have not determined if the TLS model allows for this, but at face value it seems to introduce a complication that may be easier explained in some other way.

Finally, we have observed that the IMD artifacts appear highly similar from cooldown to cooldown. Location, frequency, and other characteristics of TLS in a sample tend to change considerably from cooldown to cooldown \cite{shalibo_lifetime_2010}. No such change is seen in our experiment over a wide collection of operating points. If these artifacts are due to TLS, it would require that some collection of TLS exhibit nearly identical behavior from cooldown to cooldown. 

While we lack a comprehensive alternative explanation for these artifacts, we believe that it is more likely that they are due to higher-order nonlinearities of the SPA, which tend to dominate at higher powers as the photon count in the resonator increases. This assumption frees us to fit IMD behavior to the lowest source power regions of the measurement, which will be most similar to the powers involved in a squeezing measurement. 

Existing IMD measurement theory for SPA $K$ \cite{spa2} assumes $\Delta=0$; accounting for non-zero $\Delta$, a derivation detailed in appendix \ref{app:iip3} results in the expression
\rev{\begin{align}
    IIP_3&\approx \frac{\kappa_{\mathrm{ext}}^2 \hbar \omega_0}{|K|} \frac{1}{|\sqrt{G_{\mathrm{IL}}}e^{i\theta_g}+1|^3}\label{eq:iip3_approx_mine}
\end{align}}
Here, we use $G_{\mathrm{IL}}=|S_{11, \mathrm{on}}|^2$, the gain accounting for insertion loss of the SPA, rather than $G=\frac{|S_{11, \mathrm{on}}|^2}{|S_{11, \mathrm{off}}|^2}$. Experimentally, this requires calibration of the microwave line insertion loss through a measurement with the SPA flux-tuned far from the frequency of interest.

The angle $\theta_g$ is $\angle S_{11}$ at the input of the SPA. This can be measured with a VNA, provided that the phase shift due to cables and other microwave components is compensated. Near resonance with the amplifier off, $\angle S_{11} =0$; the phase shift due to components at that frequency can thus be measured exactly, and subtracted from the measured phase at the same frequency with the amplifier on to determine $\theta_g$. According to theory,

\begin{align}
    \theta_g=\arg \left[ i\kappa_{\mathrm{ext}} \frac{-\omega - \Delta_{\mathrm{eff}} - i\kappa/2}{\Delta_{\mathrm{eff}}^2 - |g_{\mathrm{eff}}|^2 - (\omega + i\kappa/2)^2} - 1 \right]
\end{align}


\rev{With the above expressions and the necessary VNA measurements, we are able to calculate the $|K|$ values in Fig.~\ref{fig:imd_squeezing}\subref{fig:s_kerr} from the $IIP_3$ data of Fig.~\ref{fig:imd_squeezing}\subref{fig:iip3}. See Appendix \ref{app:iip3} for additional theoretical details. 
The uncertainty in $|K|$ follows from Eq.~\ref{eq:iip3_approx_mine} through the uncertainty in $\kappa_\mathrm{ext}$, $IIP_3$, and the factor $|\sqrt{G_\mathrm{IL}}e^{i\theta_g}+1|$. Taking conservative uncertainties of $0.5$~dB on $G_\mathrm{IL}$, $0.5$~dB on $IIP_3$, $3\%$ on $\kappa_\mathrm{ext}$ from the resonator fits, and $10^\circ$ on the measured $\theta_g$, we calculate a fractional uncertainty in $|K|$ of less than 25\%. This uncertainty is low enough to not affect the conclusion of a weak dependence of $S$ on $|K|$. Separately, the uncertainty in the gain of the output line is estimated to be smaller than 3~dB, which shifts all estimated $|K|$ values in common by at most a factor of 2 without affecting their relative magnitudes. As illustrated by Figs.~\ref{fig:imd}(b) and (c), operating points that exhibited unexpected features in $IIP_3$ measurements tended to have significantly lower $IIP_3$ values and therefore higher $|K|$ values that placed them outside of the main cluster of points in Fig.~\ref{fig:imd_squeezing}(\subref{fig:s_kerr}).}

\section{IMD THEORY FOR A DETUNED SPA}
\label{app:iip3}
\input{IMDTheoryAppendix}

\section{SPA PERFORMANCE AT HIGHER GAIN}\label{app:highgain}
\begin{figure}[tp]
\centering
\begin{subfigure}{0.48\textwidth}
    \stepcounter{subfigure}
    \begin{tikzpicture}
        \node[inner sep=0pt] (img) {\includegraphics{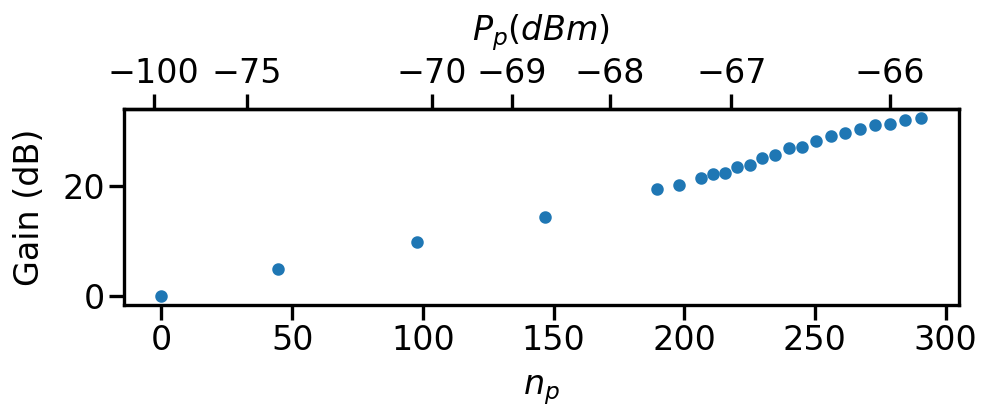}};
        \node[anchor=north west, fill=white, opacity=0.7, text opacity=1, font=\bfseries, xshift=0pt, yshift=0pt] 
            at (img.north west) {(\thesubfigure)};
    \end{tikzpicture}
    \label{fig:fine_gain_as}
\end{subfigure}

\begin{subfigure}{0.48\textwidth}
    \stepcounter{subfigure}
    \begin{tikzpicture}
        \node[inner sep=0pt] (img) {\includegraphics{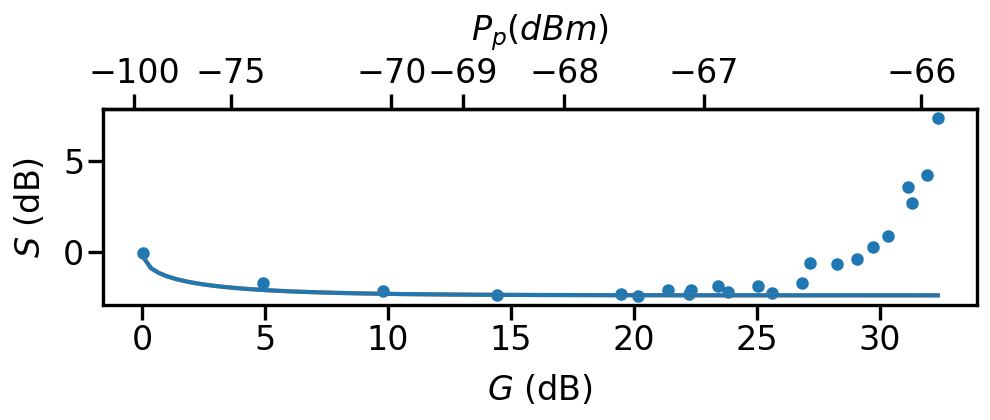}};
        \node[anchor=north west, fill=white, opacity=0.7, text opacity=1, font=\bfseries, xshift=0pt, yshift=0pt] 
            at (img.north west) {(\thesubfigure)};
    \end{tikzpicture}
    \label{fig:fine_squeezing}
\end{subfigure}

\begin{subfigure}{0.48\textwidth}
    \stepcounter{subfigure}
    \begin{tikzpicture}
        \node[inner sep=0pt] (img) {\includegraphics{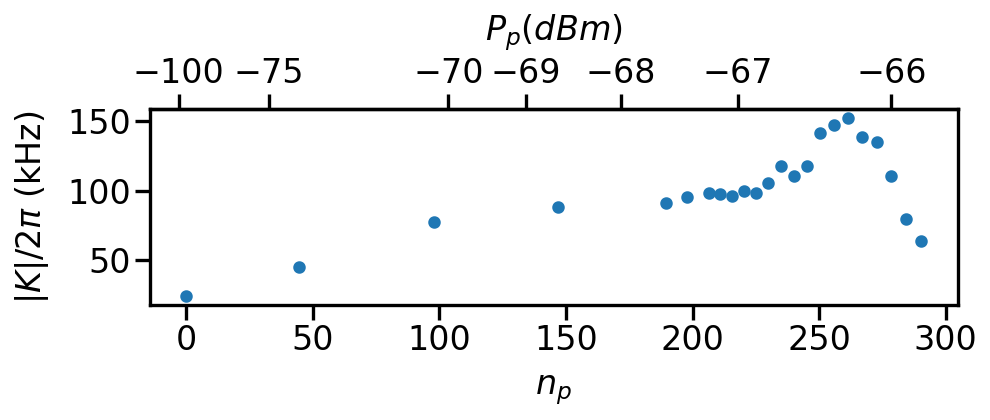}};
        \node[anchor=north west, fill=white, opacity=0.7, text opacity=1, font=\bfseries, xshift=0pt, yshift=0pt] 
            at (img.north west) {(\thesubfigure)};
    \end{tikzpicture}
    \label{fig:fine_kerr}
\end{subfigure}


\caption{\label{fig:fine_gain} (\textsubcap{\subref{fig:fine_gain_as}}) VNA-measured SPA gain over a finer set of pump powers at $\Phi_{\mathrm{ext}}=0.42\Phi_0$.  (\textsubcap{\subref{fig:fine_squeezing}}) At the same operating points, squeezing shows saturation by $G\approx10$ dB, followed by an unwanted increase in variance above the vacuum level. The solid line shows the theoretical squeezing for a given gain, and does not exhibit degradation of squeezing at high gain.  (\textsubcap{\subref{fig:fine_kerr}}) Kerr nonlinearity as measured through IMD. }

\end{figure}

The target gains of 10, 12.5, 15, and 17.5 dB were selected based on the assumption that squeezing performance would degrade as the amplifier approached saturation. But in the main measurement, the data showed squeezing that was relatively constant with respect to SPA gain. It is therefore natural to question what happens when the SPA gain is increased beyond the 17.5 dB selected as a limit in this experiment. To investigate, a measurement was performed at a flux $\Phi_{\mathrm{ext}}/\Phi_0=0.42$ with finer resolution in pump power. The results are shown in figure \ref{fig:fine_gain}. 

In this measurement, gain increases monotonically with pump power, or $n_p$, which is calculated from the equations
\begin{align}
u_p &= (\omega_p-\omega_0+\frac{2}{3}i\kappa_p)\alpha_p - \frac{K}{12}\left(\frac{32}{9}|\alpha_p|^2 \right) \alpha_p \\
P_{p, \mathrm{RT}}&=C\frac{\hbar \omega_0^3 |u_p|^2}{\kappa_p \omega_p^2}.
\end{align}

These equations are based on Eq.~(B18) of Ref.~\cite{spa1} assuming zero signal and idler power, and from input-output theory. The symbol $u_p$ is the drive strength at the pump frequency, and $P_{p, \mathrm{RT}}$ is the drive power at room temperature. The factor $C$ accounts for the attenuation of the pump drive line and was determined by a combination of room temperature line measurements and matching the measured $g_3$ in Fig.~\ref{fig:spa_setup}(\subref{fig:freqs}) to the estimate from theory.


At low $G$, $S$ conformed closely to the model presented in Appendix \ref{app:spa-squeezing}, using $\eta_{\mathrm{int}}\eta_{\mathrm{cold}}=0.42$.
However, squeezing did not noticeably improve with higher power, and eventually the variance in the squeezed quadrature became greater than the vacuum variance, yielding $S>0$. At high $G$, deviations from the model are attributed to distortion of the output state from higher-order nonlinear corrections or soft pump corrections~\cite{boutinEffectHigherorderNonlinearities2017}.


At $n_p=0$, $|K|=12|g_4^*|$. As $n_p$ increased to about 100, $|K|$ appeared to increase linearly. However, for $100<n_p<250$, $|K|$ is approximately constant, contrary to the prediction of Eq.~\eqref{eq:kerr}. Only with $n_p > 250$ does $|K|$ vary significantly. This behavior suggests that Eq.~\ref{eq:kerr}  is insufficient in realistic experimental conditions, and still higher order terms must be considered. 

\bibliography{MyLibrary}

\end{document}

%% file: squeezingTheoryAppendix.tex

Here, we develop a linearized, pump-biased description of quadrature squeezing in a single-mode SPA, tailored to the degenerate limit where $\omega_s=\omega_p/2$. Starting from the quantum Langevin equation (QLE) for the SNAIL Hamiltonian, we separate the semiclassical pump from quantum fluctuations and perform first-order harmonic balancing to retain near-resonant harmonics apropos of the signal band. The upshot is a compact susceptibility matrix that captures the effective pump-dressed detuning and parametric coupling responsible for phase-sensitive amplification and squeezing. We first analyze an ideal, lossless device in which the coupling to the measurement port equals the total linewidth $\kappa$, and then treat the more realistic case with internal dissipation to quantify how lossy channels diminish the observable squeezing. The analysis assumes a stiff pump and weak Gaussian fluctuations.

\emph{Lossless, Ideal SNAIL:} Following Ref. \cite{spa1}, the single-mode SNAIL parametric amplifier obeys the QLE
\begin{align}
\label{eq:spa-qle-full}
\dot{\hat a}
&= -\,i\omega_0\, \hat a
-\frac{\kappa}{2}\!\left(\hat a-\hat a^\dagger\right)
-3i g_3 \!\left(\hat a+\hat a^\dagger\right)^2\notag\\
&\quad-4i g_4 \!\left(\hat a+\hat a^\dagger\right)^3
+ \sqrt{\kappa}\,\hat a_{\mathrm{in}},
\end{align}
which already includes the effect of counter-rotating corrections, necessitated by the participation of out-of-band harmonics. We linearize the dynamics about a strong coherent pump amplitude and retain near-resonant terms via harmonic balancing to study fluctuations in the signal band. We decompose the field into a classical pump envelope and quantum fluctuations, $\hat a(t)=\alpha(t)+\delta\hat a(t)$, wherein the mean field $\expval{\hat{a}(t)}=\alpha(t)$ can be approximated by an undepleted tone at $\omega_p$, $\alpha(t)\approx \alpha_p\,e^{-i\omega_p t}-\Lambda_p\,\alpha_p^*\,e^{+i\omega_p t}$, with $\Lambda_p={(\omega_p-\omega_0)}/{(\omega_p+\omega_0)}$ being a dimensionless mixing weight on the conjugate tone, as introduced in \cite{spa1}. Since the resonator frequency $\omega_0$ remains in the vicinity of the half-pump tone $\omega_p/2$, we can set $\Lambda_p\approx{1}/{3}$. Finally, we expand the quantum fluctuations as $\delta\hat a(t)=\sum_{\omega}\!\delta\hat a_\omega(t)e^{-i\omega t}$. 

Substituting the above expansions into equation \eqref{eq:spa-qle-full} and truncating to first order in fluctuations separates the mean-field pump equation (used to determine the semiclassical response, e.g., gain) from the fluctuation dynamics. Isolating the harmonics oscillating at $\omega_s$ yields the near-resonant signal-band equation in drift-matrix form:
\begin{widetext}
\begin{align}
\label{eq:drift-matrix}
\frac{d}{dt}
\begin{pmatrix}
\delta\hat a_s(t)\\[2pt]
\delta\hat a_s^{\dagger}(t)
\end{pmatrix}
&=
\begin{pmatrix}
 -i\Delta_{\mathrm{eff}}-\frac{\kappa}{2} & -\,i\,g_{\mathrm{eff}}\\[4pt]
 i\,g_{\mathrm{eff}} & \,i\Delta_{\mathrm{eff}}-\frac{\kappa}{2}
\end{pmatrix}
\begin{pmatrix}
\delta\hat a_s(t)\\[2pt]
\delta\hat a_s^{\dagger}(t)
\end{pmatrix}
+\sqrt{\kappa}\,
\begin{pmatrix}
\hat a_{s,\mathrm{in}}(t)\\[2pt]
\hat a_{s,\mathrm{in}}^{\dagger}(t)
\end{pmatrix}.
\end{align}
\end{widetext}
Here $N\equiv\begin{pmatrix}
 -i\Delta_{\mathrm{eff}}-{\kappa}/{2} & -\,i\,g_{\mathrm{eff}}\\[4pt]
 i\,g_{\mathrm{eff}} & \,i\Delta_{\mathrm{eff}}-{\kappa}/{2}
\end{pmatrix}$ is the drift matrix, with pump-dressed parameters $\Delta_{\mathrm{eff}}
=\Delta+\frac{32 g_4}{3}|\alpha_p|^2$ and $g_{\mathrm{eff}}
=4 |g_3|\alpha_p$, defined in the preceding appendix. Since the pump phase merely rotates the squeezing axis, we take $\alpha_p$ to be real for convenience. The precise numerical prefactors underlying the $g_3$- and $g_4$-dependent contributions in $N$ reflect the combinatorics of three- and four-wave mixing pathways, within a first-order harmonic-balancing approximation. Since the signal tone is degenerate with the pump subharmonic ($\omega_s=\omega_p/2$), this linearized description reduces to that of a detuned degenerate parametric amplifier (DPA).

Before proceeding further, we note that higher-order harmonics and pump-induced corrections can complicate the fluctuation dynamics by renormalizing the effective strengths of both three-wave and four-wave mixing, in addition to introducing excess sideband noise. Nevertheless, as in the semiclassical case, the leading-order drift matrix is well approximated by replacing the bare \(g_4\) with \(K/12\), where \(K\) is the effective, pump-dependent Kerr coefficient estimated from $IIP_3$. A systematic, higher-order treatment would incorporate out-of-band mixing products self-consistently with their associated input fluctuations---for instance, via Floquet perturbation theory \cite{garcia-mataEffectiveFloquetTheory2024, peng_floquet-mode_2022}---which lies beyond the scope of the present model. In what follows, we focus principally on the pump subharmonic at $\omega_s=\omega_p/2$ and neglect subleading sideband-noise contributions. Hence, rather than conducting a full-fledged nonlinear analysis incorporating fluctuations, we subsume these higher-order effects at leading order by adopting the substitution $g_4\approx K/12$ in the drift matrix.


Adopting the Fourier convention $X(\Omega)\equiv\int\dd te^{i\Omega t}X(t)$ (with $X^{\dagger}(\Omega)\equiv[X(\Omega)]^{\dagger}$), and defining the fluctuation and input vectors
$\mathbf{v}(\Omega)\!\equiv\!\bigl(\delta\hat a_s(\Omega),\,\delta\hat a_s^\dagger(-\Omega)\bigr)^{T}$ and \\$\mathbf{v}_{\mathrm{in}}(\Omega)\equiv\!\bigl(\delta\hat a_{s,\text{in}}(\Omega),\,\delta\hat a_{s,\text{in}}^\dagger(-\Omega)\bigr)^{T}$,  Eq.~\eqref{eq:drift-matrix} translates into $\mathbf{v}(\Omega)
=-\sqrt{\kappa}\left(i\Omega\,\mathds{1}+N\right)^{-1}\!\mathbf{v}_{\mathrm{in}}(\Omega)$ in the frequency domain, with $\mathds{1}$ symbolizing the $2\times 2$ identity matrix. Next, by employing the input--output relation $\hat a_{s,\mathrm{out}}(\Omega)=\hat a_{s,\mathrm{in}}(\Omega)-\sqrt{\kappa}\,\delta\hat a_s(\Omega)$ and similarly defining the output vector $\mathbf{v}_{\mathrm{out}}(\Omega)\equiv\!\bigl(\delta\hat a_{s,\text{out}}(\Omega),\,\delta\hat a_{s,\text{out}}^\dagger(-\Omega)\bigr)^{T}$, the port mapping can be expressed succinctly as $\mathbf{v}_{\mathrm{out}}(\Omega)=\mathbb{S}(\Omega)\mathbf{v}_{\mathrm{in}}(\Omega)$, with the scattering matrix related to the drift matrix as
\begin{align}
    \mathbb{S}(\Omega)\equiv-\kappa\left(i\Omega\,\mathds{1}+N\right)^{-1}-\mathds{1},
\end{align}
and obeying the Bogoliubov (symplectic) identity 
\begin{align}
    \label{eq:symplectic:lossless}
    \abs{\mathbb{S}_{11}(\Omega)}^2-\abs{\mathbb{S}_{12}(\Omega)}^2=1.
\end{align}
For the output quadrature at angle $\theta$, $\hat X_\theta(\Omega)\equiv \!\left(
e^{-i\theta}\hat a_{s,\mathrm{out}}(\Omega)+
e^{+i\theta}\hat a_{s,\mathrm{out}}^\dagger(\Omega)\right)/\sqrt{2}$, the symmetrized squeezing spectrum is defined by
\begin{align}
    \mathcal{S}(\theta; \Omega) = \frac{1}{4\pi} \int d\Omega' \left\langle \{\hat{X}_\theta(\Omega), \hat{X}_\theta(\Omega')\}\right\rangle.
\end{align}
Utilizing the property of vacuum input correlations, $\expval{\delta\hat{a}_{\text{s,in}}(\Omega) \delta\hat{a}_{\text{s,in}}^{\dagger}(\Omega')}=2\pi\delta(\Omega-\Omega')$, the spectrum reduces to
\begin{align}
    \mathcal{S}(\theta;\Omega)=\frac{1}{4}\bigg[&\abs{\mathbb{S}_{11}(\Omega)e^{-i\theta}+\mathbb{S}_{12}^*(-\Omega)e^{i\theta}}^2\notag\\ &+\abs{\mathbb{S}_{12}(\Omega)e^{-i\theta}+\mathbb{S}_{11}^*(-\Omega)e^{i\theta}}^2\bigg].
\end{align}
To quantify the baseband squeezing in the signal mode, we evaluate the symmetrized spectrum at \(\Omega=0\),
\begin{align}
    \mathcal{S}(\theta)=\tfrac{1}{2}\,\bigl|\mathbb{S}_{11}e^{-i\theta}+\mathbb{S}_{12}^*e^{i\theta}\bigr|^2,
\end{align}
where \(\mathbb{S}_{ij}\equiv \mathbb{S}_{ij}(0)\) for brevity. Converted to dB by referencing the vacuum noise floor \((\mathcal{S}=1/2)\), this implies a squeezing level of $20\log_{10}\,\bigl|\mathbb{S}_{11}e^{-i\theta}+\mathbb{S}_{12}^*e^{i\theta}\bigr|$.

\emph{Finite internal loss:} Including internal dissipation introduces an additional (unmonitored) noise channel in accordance with the Langevin prescription. The linearized QLEs for the signal mode now read
\begin{widetext}
\begingroup
\setlength{\arraycolsep}{6pt}
\renewcommand{\arraystretch}{1.1}
\begin{align}
\label{eq:lossy-langevin}
\frac{d}{dt}
\begin{pmatrix}
\delta\hat a_s(t)\\[2pt]
\delta\hat a_s^{\dagger}(t)
\end{pmatrix}
=
\begin{pmatrix}
 -i\Delta_{\mathrm{eff}}-\tfrac{\kappa}{2} & -\,i\,g_{\mathrm{eff}}\\[4pt]
 i\,g_{\mathrm{eff}} & \,i\Delta_{\mathrm{eff}}-\tfrac{\kappa}{2}
\end{pmatrix}
\begin{pmatrix}
\delta\hat a_s(t)\\[2pt]
\delta\hat a_s^{\dagger}(t)
\end{pmatrix}\;+\;\sqrt{\kappa_{\mathrm{ext}}}\,
\begin{pmatrix}
\hat a_{s,\mathrm{in}}(t)\\[2pt]
\hat a_{s,\mathrm{in}}^{\dagger}(t)
\end{pmatrix}
\;+\;
\sqrt{\kappa_{\mathrm{int}}}\,
\begin{pmatrix}
\hat a_{s,\mathrm{in}}^{\prime}(t)\\[2pt]
\hat a_{s,\mathrm{in}}^{\prime\dagger}(t)
\end{pmatrix}.
\end{align}
\endgroup
\end{widetext}
Here, $\hat a_{s,\mathrm{in}}$ and $\hat a'_{s,\mathrm{in}}$ denote the external and internal (unmonitored) input fields, respectively. Fourier transforming Eq.~\eqref{eq:lossy-langevin} and using the input--output relation for the measurement port,
$\hat a_{s,\mathrm{out}}(\Omega)=\hat a_{s,\mathrm{in}}(\Omega)-\sqrt{\kappa_{\mathrm{ext}}}\,\delta\hat a_s(\Omega)$, the two-port mapping reads $\mathbf{v}_{\mathrm{out}}(\Omega)=\mathbb{S}_{\text{ext}}(\Omega)\mathbf{v}_{\mathrm{in}}(\Omega)+\mathbb{S}_{\text{int}}(\Omega)\mathbf{v}_{\mathrm{in}}'(\Omega)$, with the port-resolved scattering matrices $\mathbb{S}_{\text{ext}}(\Omega)\equiv-\kappa_{\mathrm{ext}}\left(i\Omega\,\mathds{1}+N\right)^{-1}-\mathds{1}$ and $\mathbb{S}_{\text{int}}(\Omega)\equiv-\sqrt{\kappa_{\mathrm{ext}}\kappa_{\mathrm{int}}}\left(i\Omega\,\mathds{1}+N\right)^{-1}$, which satisfy the generalized Bogoliubov identity 
\begin{align}
\sum_{\mu=\text{ext,int}}\big(\abs{\mathbb{S}_{\mu,11}(\Omega)}^2-\abs{\mathbb{S}_{\mu,12}(\Omega)}^2\big)=1.
\end{align}
Assuming uncorrelated vacuum inputs at the two ports, the zero-frequency squeezing spectrum morphs into 
\begin{align}
    \mathcal{S}_{\mathrm{lossy}}(\theta)=\frac{1}{2}\sum_{\mu=\text{ext,int}}\abs{\mathbb{S}_{\mu,11}e^{-i\theta}+\mathbb{S}_{\mu,12}^*e^{i\theta}}^2.
    \label{eq:S-lossy}
\end{align}
Equation~\eqref{eq:S-lossy} shows how internal dissipation admixes unsqueezed vacuum from the unmonitored port, thereby degrading the observable squeezing according to the coupling ratio $\kappa_{\mathrm{ext}}/\kappa$.

\emph{Relation between gain and optimal squeezing:} In a self-consistent linear model, with the dynamics linearized around a fixed operating point, both gain and squeezing are governed by the same intracavity susceptibility (often pump-dressed). The principal distinction in the fluctuation analysis stems from the extrinsic noise statistics at the ports. In a single-port, lossless device referenced to unit transmission, the (power) gain is $G=G_{\mathrm{IL}}=\abs{\mathbb{S}_{11}}^2$, where $\mathbb{S}_{11}$ is the signal transfer coefficient. With internal loss, the experimentally accessible gain at the measurement port is given by $G_{\mathrm{IL}}=\abs{\mathbb{S}_{\text{ext},11}}^2$. 

Using the symplectic constraint $|\mathbb{S}_{11}|^2-|\mathbb{S}_{12}|^2=1$ in the lossless case, and optimizing the output quadrature variance over the angle $\theta$, the minimum squeezing at zero frequency is
\begin{align}\label{eq:smin}
     \mathcal{S}_{\text{min}}
    &=\frac{1}{2}\bigg[|\mathbb{S}_{11}|^2+|\mathbb{S}_{12}|^2-2|\mathbb{S}_{11}\mathbb{S}_{12}|\bigg]\notag\\
    &=\frac{1}{2}\bigg[\sqrt{G_{\mathrm{IL}}}-\sqrt{G_{\mathrm{IL}}-1}\bigg]^2,
\end{align}
attained at the optimal quadrature orientation $\theta_{\min} = ({1}/{2})\,\arg\!\big(\mathbb{S}_{11}\mathbb{S}_{12}\big) + {\pi}/{2}$. Writing $|\mathbb{S}_{11}|=\cosh r$ and $|\mathbb{S}_{12}|=\sinh r$ with a real squeezing parameter $r\ge 0$ gives $G_{\mathrm{IL}}=\cosh^2 r$ and $\mathcal{S}_{\min}=e^{-2r}/2$, making explicit the one-to-one correspondence between gain and optimal squeezing in the absence of internal loss (i.e., $\mathcal{S}_{\min}$ is invariant along a fixed-$G_{\mathrm{IL}}$ contour).

When internal dissipation is present, the output field contains noise admixed from the unmonitored port(s). Minimizing over $\theta$ yields
\begin{equation}
\mathcal{S}_{\min,\mathrm{lossy}}
= \frac{1}{2} + \sum_{\mu} |\mathbb{S}_{\mu,12}|^{2}
- \left|\sum_{\mu} \mathbb{S}_{\mu,11}\,\mathbb{S}_{\mu,12}\right|
\label{eq:Smin-lossy}
\end{equation}
at $\theta_{\min} = ({1}/{2})\,\arg\!\Big(\sum_{\mu} \mathbb{S}_{\mu,11}\mathbb{S}_{\mu,12}\Big) + {\pi}/{2}$, where $\mathbb{S}_{\mu}$ denotes the port-resolved Bogoliubov block evaluated at zero analysis frequency, and $\mu$ indexes the external (measurement) and internal (unmonitored) channels. Owing to multiport noise interference, $\mathcal{S}_{\min,\mathrm{lossy}}$ cannot be expressed solely in terms of the measured power gain $G_{\mathrm{IL}}=|\mathbb{S}_{\mathrm{ext},11}|^2$. Consequently, the gain–squeezing mapping ceases to be one-to-one, notwithstanding their dependence on a single susceptibility. Table~\ref{table:1} summarizes the phase-preserving gain and optimal squeezing, providing their exact analytical expressions with and without internal losses.

\setlength{\extrarowheight}{2pt} 
\renewcommand{\arraystretch}{1.8} 

\newcolumntype{C}{>{\centering\arraybackslash}X} 

\begin{table*}[!htbp]
\centering
\scriptsize
\setlength{\tabcolsep}{4pt}
\begin{tabularx}{\textwidth}{@{}>{\centering\arraybackslash}p{0.22\textwidth}|C|C@{}}
\hline
 & \textbf{Lossless ($\kappa_{\mathrm{int}}=0$)} &
 \textbf{Lossy ($\kappa_{\mathrm{int}}>0$)} \tabularnewline
\hline

Symplectic constraint &
\(\,|S_{11}|^{2}-|S_{12}|^{2}=1\,\) &
\(\,\sum_{\mu=\mathrm{ext},\mathrm{int}}
\bigl(|S_{\mu,11}|^{2}-|S_{\mu,12}|^{2}\bigr)=1\,\)
\tabularnewline
\hline

$G_{\mathrm{IL}}$ &
\(\,|S_{11}|^{2}=1+{\kappa^{2} g_{\mathrm{eff}}^{2}}/{D_{0}^{2}}\,\) &
\(\,|S_{\mathrm{ext},11}|^{2}=
\bigl(\bigl(D_{0}-{\kappa_{\mathrm{ext}}\kappa}/{2}\bigr)^{2}
+(\kappa_{\mathrm{ext}}\Delta_{\mathrm{eff}})^{2}\bigr)/{D_{0}^{2}}\,\)
\tabularnewline
\hline

\(\mathcal{S}_{\min}\) &
\(\,\left(
\dfrac{\sqrt{\bigl({\kappa^{2}}/{2}-D_{0}\bigr)^{2}+(\kappa\Delta_{\mathrm{eff}})^{2}}
-\kappa g_{\mathrm{eff}}}{\sqrt{2}\,D_{0}}
\right)^{2}\,\)&
\(\,{1}/{2}
+{\kappa_{\mathrm{ext}}\kappa\,g_{\mathrm{eff}}^{2}}/{D_{0}^{2}}
-\frac{({\kappa_{\mathrm{ext}} g_{\mathrm{eff}}})\,
\sqrt{\bigl({\kappa^{2}}/{2}-D_{0}\bigr)^{2}+(\kappa\Delta_{\mathrm{eff}})^{2}}}{{D_{0}^{2}}}\,\)
\tabularnewline
\hline
\end{tabularx}
\caption{Lossless versus lossy operation: phase-preserving gain and optimal squeezing. Here \(D_{0}=\Delta_{\mathrm{eff}}^{2}+(\kappa/2)^{2}-g_{\mathrm{eff}}^{2}\), with \(g_{\mathrm{eff}}\) taken real and nonnegative. The expressions are exact within the linearized model and remain meaningful insofar as the stability criterion \(D_{0}>0\) is met; $D_0(n_p=0)\geq 0$, so violating this criterion would require increasing the pump power beyond the parametric instability point, an atypical operating regime \cite{malnouOptimalOperationJosephson2018}}.
\label{table:1}
\end{table*}

\emph{Effect of attenuated transmission:} Besides internal-loss channels, squeezing is further degraded by finite transmission efficiency in the measurement chain. One typically models downstream losses using a beam-splitter description; i.e., the detected field is an admixture of the device output and vacuum noise. For an external power transmission efficiency $\eta$, the squeezing observed at the detector plane would be obtained as
\begin{widetext}
\begin{align}
\label{eq:Sobs}
\mathcal S_{\mathrm{obs}}
= \eta\,\mathcal S_{\min,\mathrm{lossy}} + (1-\eta)\,\frac{1}{2}= \frac{1}{2}
+ \frac{(\eta\,\kappa_{\mathrm{ext}})\,\kappa\,g_{\mathrm{eff}}^{2}}{D_{0}^{2}}
- \frac{(\eta\,\kappa_{\mathrm{ext}})\,g_{\mathrm{eff}}}{D_{0}^{2}}
  \sqrt{\Bigl(\frac{\kappa^{2}}{2}-D_{0}\Bigr)^{2}+(\kappa\Delta_{\mathrm{eff}})^{2}}\,.
\end{align}
\end{widetext}
where $S_{\min,\text{lossy}}$ and $D_0$ are both defined in Table \ref{table:1}. Equivalently, the two loss mechanisms enter as an effective reduction of the external coupling rate, ${\kappa_{\mathrm{ext}}\rightarrow \eta\kappa_{\mathrm{ext}}}$, in the expression for the observed squeezing. This becomes particularly transparent when we set $\Delta_{\mathrm{eff}}\approx 0$. In this limit, Eq.~\eqref{eq:Sobs} considerably simplifies to:
\begin{align}
    \mathcal S_{\mathrm{obs}}=\frac{1}{2}-\frac{(\eta\kappa_{\mathrm{ext}})\,g_{\text{eff}}}{(g_{\text{eff}}+\kappa/2)^2}
\end{align}
As an illustration of the impact of these nonidealities, consider representative parameters close to the experimental values: $\kappa/2\pi=340$ MHz, $g_3/2\pi=2$ MHz, and $\alpha_p=11.5$. We further set $\Delta_{\mathrm{eff}}\approx 0$.
\begin{itemize}
  \item With $\kappa_{\mathrm{int}}=0$ and perfect power transmission $\eta = 1$, the predicted squeezing is {$-10.5$ dB}. Introducing modest internal loss, e.g., $\kappa_{\mathrm{ext}}=0.9\,\kappa$ ($\kappa_{\mathrm{int}}=0.1\kappa$), the squeezing drops precipitously to {$-7.5$ dB}.
  \item In practice, both $\eta$ and $\kappa_{\mathrm{int}}$ limit the observed squeezing. For example, $\kappa_{\mathrm{ext}}=0.8\,\kappa$ (i.e., $\kappa_{\mathrm{int}}\approx 0.2 \kappa$) and $\eta=0.60$ yields $-2.5$~dB of observed squeezing.
\end{itemize}

These examples manifest that internal dissipation and downstream transmission losses compound, substantially depressing the observable squeezing relative to the intrinsic, device-level limit. We can define the device coupling efficiency $\eta_{\text{int}}\equiv\kappa_{\mathrm{ext}}/\kappa$, which quantifies the fraction of the intracavity field exiting through the measurement port rather than being lost internally. Then, the observable noise reduction is governed by the product of $\eta_{\text{int}}$, and the end-to-end transmission $\eta$, which accounts for insertion loss in circulators/isolators, cabling and connectors, impedance mismatch, noise added by amplifiers, and detector inefficiency. \emph{For a given linewidth $\kappa$}, a convenient aggregate figure of merit that unifies these primary loss mechanisms is the effective coupling efficiency $\eta\eta_{\text{int}}$. Consequently, improvements to $\eta$ and $\eta_{\mathrm{int}}$ are synergistic: gains in one channel can be nullified by shortfalls in the other, and the experimentally accessible squeezing is ultimately limited by the weaker of the two. 

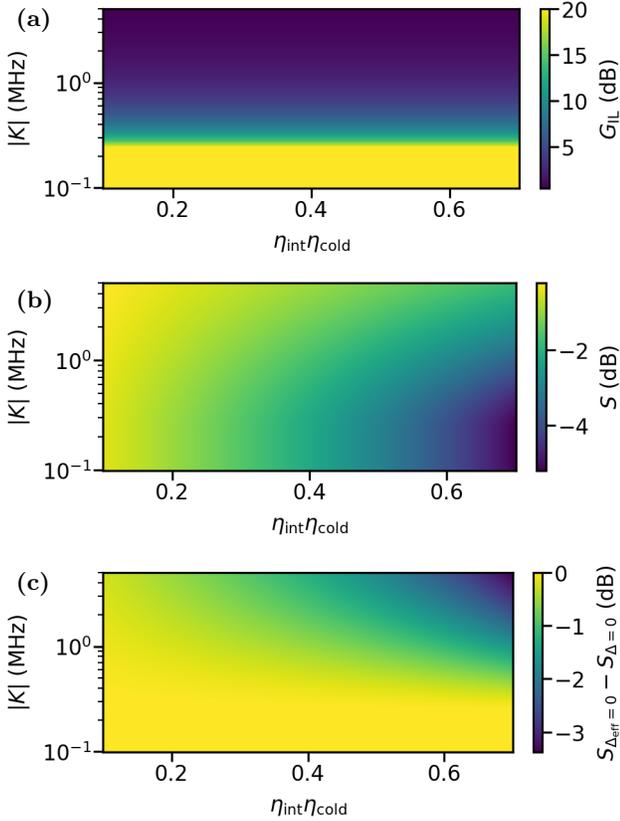
\begin{figure}[!htbp]
\centering
\begin{subfigure}{0.48\textwidth}
    \stepcounter{subfigure}
    \begin{tikzpicture}
        \node[inner sep=0pt] (img) {\includegraphics{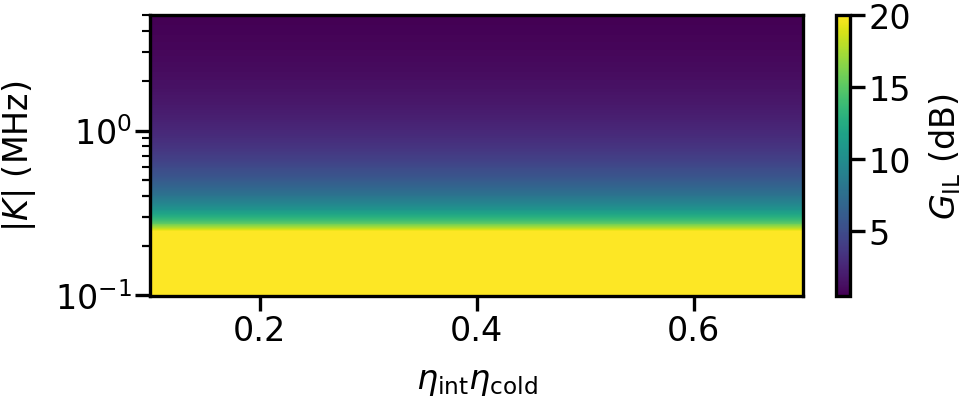}};
        \node[anchor=north west, fill=white, opacity=0.7, text opacity=1, font=\bfseries, xshift=0pt, yshift=0pt] 
            at (img.north west) {(\thesubfigure)};
    \end{tikzpicture}
    \label{fig:projection_gain}
\end{subfigure}

\vspace{10pt}

\begin{subfigure}{0.48\textwidth}
    \stepcounter{subfigure}
    \begin{tikzpicture}
        \node[inner sep=0pt] (img) {\includegraphics{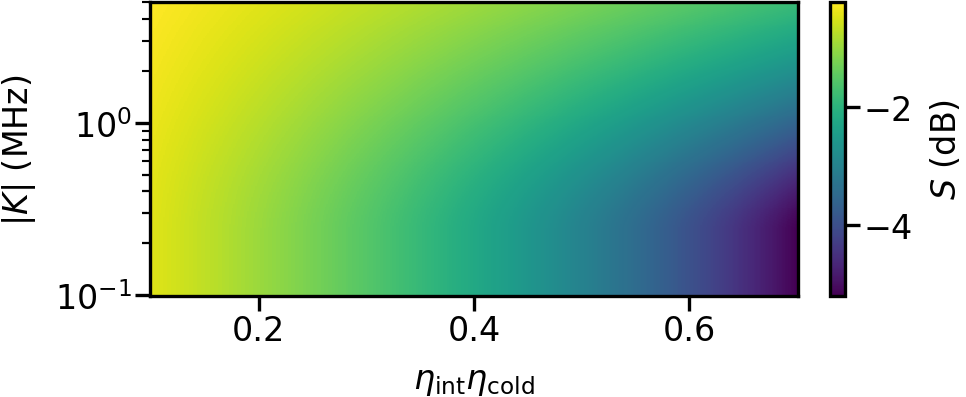}};
        \node[anchor=north west, fill=white, opacity=0.7, text opacity=1, font=\bfseries, xshift=0pt, yshift=0pt] 
            at (img.north west) {(\thesubfigure)};
    \end{tikzpicture}
    \label{fig:projection_s}
\end{subfigure}

\vspace{10pt}

\begin{subfigure}{0.48\textwidth}
    \stepcounter{subfigure}
    \begin{tikzpicture}
        \node[inner sep=0pt] (img) {\includegraphics{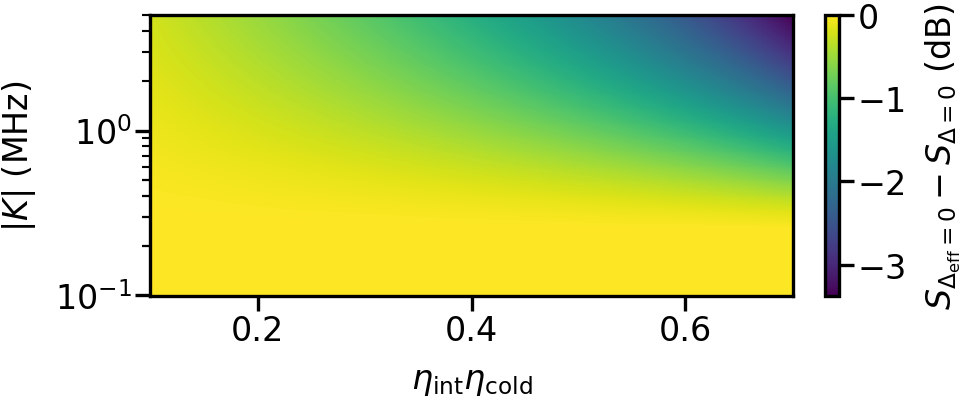}};
        \node[anchor=north west, fill=white, opacity=0.7, text opacity=1, font=\bfseries, xshift=0pt, yshift=0pt] 
            at (img.north west) {(\thesubfigure)};
    \end{tikzpicture}
    \label{fig:projection_improvement}
\end{subfigure}

\caption{\label{fig:projections} (\textsubcap{\subref{fig:projection_gain}}) Theoretical gain values of an amplifier with $g_3/2\pi=2$ MHz operated at $\Delta=0$ MHz, with $n_p$ chosen to yield $G$ as close to 20 dB as possible at each point. (\textsubcap{\subref{fig:projection_s}}) Theoretical squeezing generated by the device using the values of $n_p$ calculated for (\textsubcap{\subref{fig:projection_gain}}) to get $G$ close to 20~dB. Within the $G=20$ dB region at the bottom of (\textsubcap{\subref{fig:projection_gain}}), $K$ has no effect. As $|K|$ becomes increasingly large, the amplifier maximum gain decreases due to Stark shift and squeezing worsens. (\textsubcap{\subref{fig:projection_improvement}}) Improvement in squeezing obtained by operating the amplifier at $\Delta_{\mathrm{eff}}=0$; this allows $G=20$ dB at every $K$, $\eta_{\mathrm{int}}\eta_{\mathrm{cold}}$ point and eliminates any dependence of $S$ on $K$. In the yellow bottom of the graph, $|K|$ was too small to prevent $G=20$ dB for the simple $\Delta=0$ case, so there is no improvement. Above this region, there is an improvement in squeezing that scales with $\eta_{\mathrm{int}}\eta_{\mathrm{cold}}$.}
\end{figure}

It is also worth noting that the sole effect of Kerr nonlinearity in this model is to shift the SPA resonance; once the operating point is retuned, the optimal squeezing is not intrinsically degraded. Although this conclusion emerges within a first-order harmonic balancing framework, it accords with the empirical observation that squeezing correlates very weakly with the Kerr strength. Nonetheless, when the resonator is tuned exactly to the pump subharmonic ($\Delta=0$), the theory predicts a Kerr dependence; with high Kerr, the resonator Stark shifts faster than increasing pump power increases the gain, limiting the maximum achievable $G_{\mathrm{IL}}$ and $S$. Figure~\ref{fig:projections} illustrates this effect. With a more careful amplifier tuning to operate at $\Delta_{\mathrm{eff}}=0$, gain and therefore squeezing can be made independent of $|K|$, with squeezing degrading solely with $\eta\eta_{\mathrm{int}}$.

As a final remark, we reemphasize that higher-order sidebands---captured only in a higher-order harmonic-balancing or Floquet treatment---inject additional noise and further diminish the observable squeezing. In practice, the achievable squeezing, therefore, is unavoidably lower than the ideal DPA prediction.

%% file: IMDTheoryAppendix.tex

This appendix presents a compact derivation of third‑order (two-tone) IMD and the associated $IIP_3$ for a single-mode SPA with Kerr-type nonlinearity. The main purpose is to extend the analysis of Frattini \emph{et al.}~\cite{spa1} by highlighting the explicit role of pump--resonator detuning---an effect omitted in the earlier treatment---and by accounting for untapped, internal cavity losses. We adopt a weak two-tone excitation $\Omega_{\text{in}}(t)=\sum_{j=1}^2\Omega_{j,\text{in}}e^{-i\omega_{j}t}$ to model the drive-resonator coupling via ${\mathcal H_{\mathrm{drive}}}/{\hbar}
= \bigl(\Omega_{\mathrm{in}}+\Omega_{\mathrm{in}}^{*}\bigr)\,(\hat a+\hat a^{\dagger})$, with constituent (angular) frequencies \(\omega_{1}=2\pi f_{1}\) and \(\omega_{2}=2\pi f_{2}\) near the SPA resonance \(\omega_0=2\pi f_{r}\), and work within the standard input–output framework. Due to the Kerr-induced four-wave mixing, the nearest third-order products appear at $\omega_{\mathrm{imd},1}\equiv2\omega_{1}-\omega_{2}$ and $\omega_{\mathrm{imd},2}\equiv2\omega_{2}-\omega_{1}$,
which lie close to the fundamental tones and thus within typical measurement bandwidths. Without loss of generality, we focus below on $\omega_{\text{imd}}\equiv2\omega_{1}-\omega_{2}$, suppressing the index for brevity. We formulate our analysis under the following conditions: (i) \emph{weak nonlinearity} (operation well below gain compression), and (ii) a \emph{narrowband approximation}, in which each input tone $\omega_{j}$ lies within the amplifier bandwidth around $\omega_0$, so that the response functions (e.g., amplifier gain) and the relaxation rate (linewidth) vary slowly over the tone spacing. Further, for IIP3 extraction, we assume \emph{equal power per applied tone}, $P_{1}=P_{2}\equiv P_0$, when evaluating the intercept.

Let us denote the intracavity coherent amplitudes at the two applied tones by \(\alpha_{1}\) and \(\alpha_{2}\)  and the IMD sideband amplitude at \(\omega_\text{imd}\) by \(\alpha_\text{imd}\). Upon first-order harmonic balancing, the quantum Langevin equation (QLE) for the IMD mode reads
\begin{align}
\bigl[\omega_{\text{imd}}-\omega_0 + i\,\tfrac{\kappa}{2}\bigr] \alpha_{\text{imd}}
= & 12\,i\,g_4\,\alpha_{1}^2 \alpha_{2}^*\notag\\
&\;+\; \text{(higher-order mixing terms)},
\label{eq:qle-imd}
\end{align}
where $\kappa\equiv\kappa_{\mathrm{ext}}+\kappa_{\mathrm{int}}$ characterizes the full linewidth, including dissipative contributions from both measurement and internal-loss channels. Equivalently, the IMD tone experiences a Kerr-mediated effective input drive that scales with the strength of $g_4$,
\begin{equation}
\Omega^{\text{(eff)}}_{\text{imd,in}} \equiv 12\,  g_4\, \alpha_{1}^{2}\, \alpha_{2}^{*}.
\label{eq:cubic-law}
\end{equation}
Following Ref. \cite{spa1}, the power in an input tone at angular frequency \(\omega_{\beta}\) can be expressed as
\begin{equation}
P_{\beta,\text{in}}=\frac{\hbar\omega_0}{\kappa_{\mathrm{ext}}}\,|\Omega_{\beta,{in}}|^2\left(\frac{\omega_0}{\omega_{\beta}}\right)^2
\equiv \frac{\nu_{\beta}^2\,\hbar\omega_0}{\kappa_{\mathrm{ext}}}\,|\Omega_{\beta,\text{in}}|^2,
\label{eq:Ps}
\end{equation}
where \(\nu_{\beta}\!\equiv\!\omega_0/\omega_{\beta}\), \(\Omega_{\beta,\text{in}}\) is the complex drive amplitude, and $\beta\in\{1,2,\text{imd}\}$ labels the relevant tones proximal to the signal frequency. At this point, it is convenient to connect the input signal strength to the device response. Employing the semiclassical equation of motion for any of these tones and representing the weak-signal susceptibility as
$\chi_{\beta}$, we can write the intracavity amplitude at $\omega_{\beta}$ and the corresponding photon number as
\begin{equation}
\alpha_{\beta} = \chi_{\beta} \Omega_{\beta,\text{in}},\qquad
n_{\beta}\equiv|\alpha_{\beta}|^2=|\chi_{\beta}|^2|\Omega_{\beta,\text{in}}|^2.
\label{eq:ns}
\end{equation}
The specific form of the susceptibility is predicated on the existence of a pump tone at $\omega_p$. In the absence of a pump, the inverse of the weak-signal susceptibility takes the customary linear-response form $\chi_{\beta}^{-1}=\delta\omega_{\beta}-\Delta+i\,\kappa/2$, where \(\delta\omega_{\beta}\equiv \omega_{\beta}-\omega_p/2\) and \(\Delta\equiv \omega_0-\omega_p/2\) are defined in the frame rotating at \(\omega_p/2\). When a pump tone is switched on, i.e., $\Omega_{\text{in}}(t)\rightarrow\sum_{j=1}^2\Omega_{j,\text{in}}e^{-i\omega_{j}t}+\Omega_{p}e^{-i\omega_pt}$, with $|\Omega_{j}|\ll|\Omega_p|$, the detuning $\Delta$ incurs a pump-induced Stark shift $\Delta\rightarrow\Delta_{\text{eff}} \equiv\Delta + \frac{32g_4}{3} |\alpha_p|^2$, and the device behaves as a degenerate parametric amplifier (DPA) with an effective mixing strength $g_{\rm eff}\equiv4g_3|\alpha_p|$ (under the stiff-pump approximation). This changes the inverse susceptibility of the $\beta^{\text{th}}$ tone accordingly to $\chi_{\beta}^{-1} = (-\delta\omega_{\beta}-\Delta_{\text{eff}}-i{\kappa}/{2})^{-1}\,D_{\text{eff}}$, with $D_{\text{eff}}\equiv\Delta_\text{eff}^2 - \left(\delta\omega_{\beta} + i {\kappa}/{2} \right)^2 - g_\text{eff}^2$.

Next, invoking the input--output relation for a single mode and comparing the components oscillating at  \(\omega_{\beta}\), we obtain the output signal amplitude $\Omega_{\beta,\text{out}} =  i {\kappa_{\mathrm{ext}}} \alpha_{\beta} - \Omega_{\beta,\text{in}} = (i\kappa_{\mathrm{ext}}\chi_{\beta}-1)\Omega_{\beta,\text{in}}$. Hereafter, we drop the subscript on the susceptibility under the narrowband approximation, letting $\chi_{\beta}\approx\chi_s$ $\forall\,\beta\in\{1,2,\text{imd}\}$, and define $\delta\omega_{\beta}\equiv\omega$ for conciseness. Accordingly, identifying the generic prefactor $(i\kappa_{\mathrm{ext}}\,\chi_s-1)$ as the complex phase-preserving \emph{amplitude gain}, we conveniently parameterize it as 
\begin{align}
    \label{eq:amp-gain}
    i\,\kappa_{\mathrm{ext}}\, \chi_s - 1\;\equiv\;\sqrt{G_{\mathrm{IL}}}\,e^{i\theta_g},
\end{align}
where $G_{\mathrm{IL}}=|i\,\kappa_{\mathrm{ext}}\, \chi_s - 1|^2$ stands for the power gain referenced to unit transmission and $\theta_g=\arg(i\,\kappa_{\mathrm{ext}}\, \chi_s - 1)$ encodes the phase of the output tone relative to the input. Inserting $\chi_s=-i(\sqrt{G_{\mathrm{IL}}}e^{i\theta_g}+1)/\kappa_{\mathrm{ext}}$ into Eq.~\eqref{eq:ns}, and eliminating $|\Omega_{\beta,\text{in}}|^2$ between Eqs. \eqref{eq:Ps} and \eqref{eq:ns} yields
\begin{align}
n_{\beta} = \frac{|\sqrt{G_{\mathrm{IL}}}\,e^{i\theta_g}+1|^2}{\nu_{\beta}^2}\,\frac{P_{\beta,\text{in}}}{\hbar\omega_0\,\kappa_{\mathrm{ext}}}\\
= \frac{|\sqrt{G_{\mathrm{IL}}}\,e^{i\theta_g}+1|^2}{\nu_{\beta}^2\,G_{\mathrm{IL}}}\,\frac{P_{\beta,\mathrm{out}}}{\hbar\omega_0\,\kappa_{\mathrm{ext}}},
\label{eq:n-from-P}
\end{align}
where, in the final equality, the linear gain relation $P_{\beta,\text{in}}=P_{\beta,\mathrm{out}}/G_{\mathrm{IL}}$ has been used. Treating the Kerr-induced source term \(\Omega^{\text{(eff)}}_{\text{imd,in}}\) as a coherent input at \(\omega_\text{imd}\), and using the power mapping in Eq.~\eqref{eq:Ps}, the detected IMD output power scales as
\begin{equation}
P_{\text{imd,out}}={G_{\mathrm{IL}}}\;\frac{\hbar\omega_0}{\kappa_{\mathrm{ext}}}\;
\nu_{\text{imd}}^2\;|12 g_4|^2\; n_{1}^2 n_{2}.
\label{eq:Pimd}
\end{equation}
Substituting Eq.~\eqref{eq:n-from-P} for either of the applied tones into Eq.~\eqref{eq:Pimd} yields, in terms of the measured \emph{per-tone} outputs,
\begin{equation}
P_{\text{imd,out}}=
|12 g_4|^2\,\nu_{\text{imd}}^2\,
\frac{|\sqrt{G_{\mathrm{IL}}}e^{i\theta_g}+1|^6}{(\hbar\omega_0)^2\kappa_{\mathrm{ext}}^4}
\frac{P_{1,\text{out}}^2\,P_{2,\text{out}}}{\nu_1^4\,\nu_2^2\,G_{\mathrm{IL}}^2}.
\label{eq:Pimd-Ps}
\end{equation}
For equal per-tone output powers \(P_{1,\text{out}}=P_{2,\text{out}}\equiv P_{0,\text{out}}\), the (extrapolated) output third-order intercept follows from the standard condition $P_{\text{imd,out}}=P_{0,\text{out}}$. Solving this and referring back to the input, we obtain
\begin{equation}
\, IIP_3 
=\frac{P_{\text{imd,out}}}{G_{\mathrm{IL}}}= \Bigl(\frac{\nu_{1}^{2}\,\nu_{2}}{\nu_{\text{imd}}}\Bigr)
\, \frac{\kappa_{\mathrm{ext}}^{2}\,\hbar\omega_0}{12\,|g_4|}
\, \frac{1}{|\sqrt{G_{\mathrm{IL}}}\,e^{i\theta_g}+1|^{3}},\, 
\label{eq:IIP3-gen}
\end{equation}
In the narrowband limit near resonance, we may reasonably approximate $\nu_{\beta}\approx 1$ for each $\beta\in\{1,2,\text{imd}\}$, the above equation reducing to
\begin{equation}
\, IIP_3  \;\approx\; \frac{\kappa_{\mathrm{ext}}^{2}\,\hbar\omega_0}{12\,|g_4|}\; \frac{1}{|\sqrt{G_{\mathrm{IL}}}\,e^{i\theta_g}+1|^{3}}.\, 
\label{eq:IIP3-simple}
\end{equation}
Equation \eqref{eq:IIP3-simple} is exact within a first-order (harmonic-balancing) treatment, which neglects both higher-order harmonic generation and pump-induced super-quartic corrections. More generally, the quartic nonlinearity inferred from $IIP_3$ is best represented by an effective coefficient $g_4^{(\text{eff})}\equiv K/12$,  with $K$ (the Kerr constant) providing a more faithful measure of the net four-wave interaction inferred from (c.f. \cite{spa1,spa2}). The coefficient $K$ includes both pump-independent corrections ensuing from lower-order anharmonicities (specifically those generated by $g_3$) and pump-dependent contributions due to super-quartic mixing associated with higher-order even terms (e.g. $g_6$ and beyond). These corrections scale as \(\mathcal{O}(g_{3}^{2}/\omega_{0})\) and \(\mathcal{O}(n_{p})\) respectively. The latter is particularly relevant near flux biases where the pump-off $IIP_3$ predicts a pronounced dip in the estimated Kerr strength. Nonetheless, Eq.~\eqref{eq:IIP3-simple} remains a good approximation under the substitution \(g_{4}\to K/12\), with the proviso that \(K\) is not a fixed device constant but generally varies with pump strength.

\begin{figure}
\centering
\includegraphics{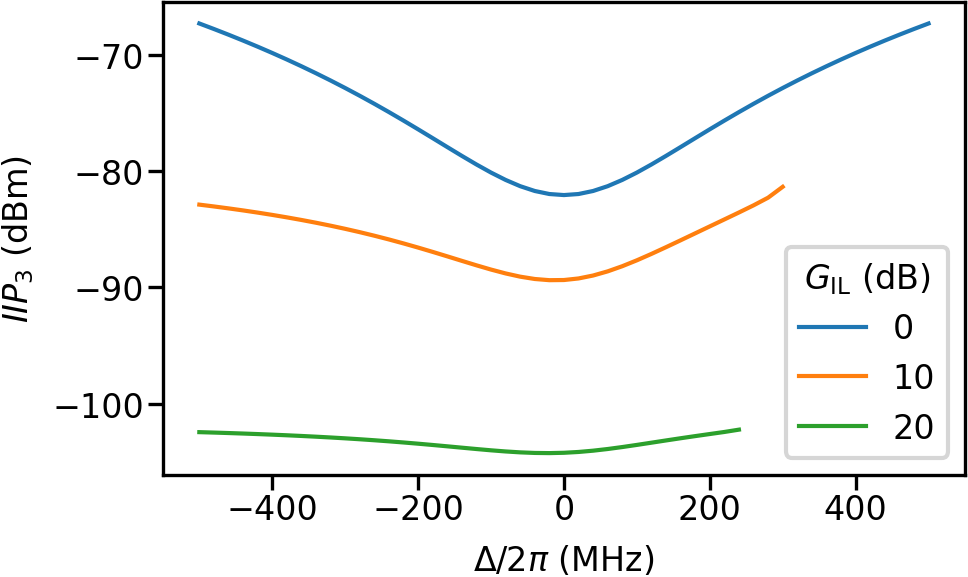}
\caption{\label{fig:iip3_delta} Theoretical $IIP_3$ values for an SPA with $\kappa=\kappa_{ext}=2\pi \cdot 340$~MHz, $g_3/2\pi=2$~MHz, and $K/2\pi=70$~kHz. For each of the three target $G_{\mathrm{IL}}$ values, $\Delta$ is swept and $n_p$ tuned to achieve the desired gain. The dependence of $IIP_3$ on $\Delta$ is strong for low gain but becomes less significant at higher gains.}
\end{figure}

For relatively low values of the power gain $G_{\mathrm{IL}}$, the detuning-dependent phase factor can substantially alter the inferred $IIP_3$ even along a constant-gain contour. Indeed, for fixed $G_{\mathrm{IL}}$, the denominator $|\sqrt{G_{\mathrm{IL}}}\,e^{i\theta_g}+1|^{3}$ can swing between the extremal values $\left( \sqrt{G_{\mathrm{IL}}} - 1 \right)^3$ and $\left( \sqrt{G_{\mathrm{IL}}} + 1 \right)^3$ as the detuning $\Delta$ is swept to rotate $\theta_g$. As an example, when $G \approx 10$ dB (also 10 in linear units), these bounds are approximately $10$ and $72$, respectively. Consequently, the upper bound can exceed the lower bound by several fold, indicating that detuning-driven swings in $IIP_3$ cannot be neglected when estimating $g_4$ (or more generally, $K$). Only in the large-gain limit \(G_{\mathrm{IL}}\gg 1\) do these bounds nearly coincide, effectively muting the detuning dependence. The asymptotic scaling then reduces to $IIP_3 \approx(\kappa_{\mathrm{ext}}^{2}\,\hbar\omega_0/|K|)\; G_{\mathrm{IL}}^{-3/2}$. The sensitivity of $IIP_3$ to pump-resonator detuning at finite amplifier gain is portrayed in Fig.~\ref{fig:iip3_delta}, with the variations largely accentuated at lower gains where $\theta_g$ plays a pivotal role. As the gain increases, this dependence progressively flattens, illustrating the predicted convergence to the asymptotic behavior.

For completeness, we note that in the absence of internal loss (i.e., $\kappa_{\mathrm{ext}}=\kappa$), setting $\Delta_{\text{eff}}=\omega=0$ reduces Eq.~\eqref{eq:IIP3-simple} to the detuning-independent, approximate $IIP_3$ expression reported in Refs. \cite{spa1,spa2}. In this limit, the phase-preserving amplitude gain $i\kappa\chi_s-1$ is purely real, implying $\theta_g=0$. Our derivation, therefore, shows that finite internal loss primarily rescales the IMD response rather than altering the cubic power law implied by Eq.~\eqref{eq:Pimd-Ps}. Alternatively stated, the same functional form for $IIP3$ applies, provided the power mapping is referenced to the measurement port, with \(G_{\mathrm{IL}}\) and \(\theta_g\) interpreted as the port-resolved gain and phase, respectively.

In closing, we note that although the semiclassical dynamics of an SPA are largely insensitive to modest internal loss, the impact of $\kappa_{\mathrm{int}}$ on quantum correlation properties---such as squeezing---can be markedly more deleterious. This is because any parasitic-loss channel feeds excess noise into the system, as detailed in appendix~\ref{app:spa-squeezing}.